\documentclass[11pt]{amsart}

\usepackage{amsmath}
\usepackage{amsthm}
\usepackage{amsfonts}
\usepackage{graphicx}
\usepackage{psfrag}

\theoremstyle{plain}
\newtheorem{thm}{Theorem}

\newtheorem{prop}{Proposition}
\newtheorem{lem}{Lemma}

\newcommand{\del}{{\partial}}
\newcommand{\DEC}{{I}}
\newcommand{\past}{{II}}
\newcommand{\rbounds}{{III}}
\newcommand{\mpos}{{IV}}
\newcommand{\drduneg}{{V}}
\newcommand{\drdvpos}{{VI}}
\newcommand{\extprinc}{{VII}}
\newcommand{\Acond}{{A}}
\newcommand{\Aprimecond}{{A}}
\newcommand{\Bcondone}{{B1}}
\newcommand{\Bcondtwo}{{B2}}
\newcommand{\Ccond}{{C}}

\begin{document}

\title[Asymptotic behavior of spherically symmetric MTTs]{Asymptotic behavior of spherically symmetric marginally trapped tubes}
\author{Catherine Williams} 
\address{Department of Mathematics, Box 354350, University of Washington, Seattle, WA 98195}
\email{cwilliam@math.washington.edu}      
\date{\today}          

\begin{abstract}We give conditions on a general stress-energy tensor $T_{\alpha \beta}$ in a spherically symmetric black hole spacetime which are sufficient to guarantee that the black hole will contain a (spherically symmetric) marginally trapped tube which is eventually achronal, connected, and asymptotic to the event horizon. 
Price law decay per se is not required for this asymptotic result, and in this general setting, such decay only implies that the marginally trapped tube has finite length with respect to the induced metric.
We do, however, impose a smallness condition (\Bcondone) which one may obtain in practice by imposing decay on the $T_{vv}$ component of the stress-energy tensor along the event horizon.  
We give two applications of the theorem to self-gravitating Higgs field spacetimes, one using weak Price law decay, the other certain strong  smallness and monotonicity assumptions.
\end{abstract}

\maketitle

\section{Introduction}
Black holes have long been one of the most-studied features of general relativity.  
The global nature of their mathematical definition, however, makes them somewhat inaccessible to physical considerations (e.g. black hole mechanics) or numerical simulation.  
In recent years, considerable work has gone into developing more tractable quasi-local notions to capture black hole behavior.  
In particular, a program initiated by Hayward and modified and refined by the work of Ashtekar and Krishnan \cite{H, A} proposes certain hypersurfaces to model the surfaces of dynamical and equilibrium black holes, called dynamical and isolated horizons respectively; such horizons are examples of a more general class of hypersurfaces known as marginally trapped tubes.  
The geometry of such hypersurfaces has been the subject of a flurry of investigation recently, and nice existence, uniqueness, and compactness results for marginally trapped tubes have been established \cite{AG, AMS, AM}.  
Major open questions remain, however. 
One area for further exploration is the relationship between marginally trapped tubes and the event horizons of black holes.
Numerical simulations appear to indicate that marginally trapped tubes that form during gravitational collapse or in the collision of two black holes should become achronal and asymptotically approach the event horizon \cite{A, SKB, B2}, but this prediction has not been proven except in two specific matter models: that of a scalar field in a (possibly) charged spacetime \cite{C2, D1}, and that of collisionless matter \cite{DR}.

In this paper, we give conditions on a \it general \rm stress-energy tensor $T_{\alpha \beta}$ in a spherically symmetric black hole spacetime which are sufficient to guarantee both that the black hole will contain a (spherically symmetric) marginally trapped tube and that that marginally trapped tube will be achronal, connected, and asymptotic to the event horizon.  We then derive some additional results pertaining to the affine lengths of both the black hole event horizon and the marginally trapped tube and show how our main result can be applied for Vaidya and Higgs field matter models.

A spherically symmetric spacetime admits an $SO(3)$-action by isometries, so it is both natural and convenient to formulate and prove these results at the level of the 1+1 Lorentzian manifold obtained by taking a quotient by this action.  
In particular, we restrict ourselves to a characteristic rectangle in Minkowski 2-space with conformal metric and past boundary data constrained in such a way that the rectangle could indeed lie inside the quotient of a spherically symmetric black hole spacetime, with one of its edges coinciding with the event horizon.  
In order to make this regime both generic and physical, we assume that our spacetime is a globally hyperbolic open subset of this characteristic rectangle and contains its two past edges; this would indeed be the case if it were the maximal future development of some initial data for the metric and stress-energy tensor prescribed along these hypersurfaces.
We use no explicit evolution equations for $T_{\alpha \beta}$.  Instead we assume that $T_{\alpha \beta}$ satisfies the dominant energy condition throughout the spacetime, and in addition we require that the spacetime satisfy a nontrivial extension principle, one which arises in the evolutionary setting for many `physically reasonable' matter models.  
Our conditions then take the form of four inequalities which must hold near a point which we call future timelike infinity and denote by $i^+$.  The inequalities relate components of the stress-energy tensor to the conformal factor and radial function for the metric.  

It is worth mentioning that the conditions we impose on $T_{\alpha \beta}$ do not directly include or imply `Price's law'.  
(Originally formulated as an estimate of the decay of radiation tails of massless scalar fields in the exterior of a black hole \cite{P}, the appellation `Price's law' is now widely used to refer to inverse power decay of any black hole ``hair" along the event horizon itself.)  
In \cite{D1}, which addressed the double characteristic initial value problem for the Einstein-Maxwell-scalar field equations, Dafermos showed that imposing a weak version of Price law decay on data along an outgoing characteristic yields a maximal future development which does indeed contain an achronal marginally trapped tube asymptotic to the event horizon.  
Consequently, one might have expected such decay to be central for obtaining the same result in the general setting.  
In this paper, however, we show that the analogous decay of $T_{vv}$ ($v$ an outgoing null coordinate) is only \it a priori \rm  related to the length of the marginally trapped tube, not its terminus.  
The conditions we use instead to control the tube's asymptotic behavior entail only smallness and integrability of certain quantities. 

However, it appears that some sort of decay is always necessary in order to retrieve our conditions in practice.  
Indeed, in the self-gravitating Higgs field setting, our conditions follow rather naturally from the assumption of weak Price-law-like decay on the derivatives of the scalar field and the potential  (Theorem \ref{HiggsR}), exactly analogously to Dafermos' result for Einstein-Maxwell-scalar fields.  
On the other hand, in Theorem \ref{HiggsM}, we are able to derive these conditions without making use of an explicit decay rate, instead using only smallness and monotonicity, and indeed one can construct examples which satisfy our conditions but violate even the weak version of Price's law. 
Still, the specific monotonicity assumptions are themselves quite strong and do imply decay, if not that which is specifically called Price's law.

The paper proceeds as follows:  in Section \ref{ba}, we present the setting for the main theorem, including all the assumptions necessary to insure that our characteristic rectangle represents the correct portion of a black hole and precise statements of the energy condition and extension principle to be used.  In Section \ref{aandc}, we present a weak version of the first of the conditions used in the main result, show in Proposition \ref{achronal} that it is sufficient by itself to guarantee achronality of the marginally trapped tube, and then use that result to establish Proposition \ref{connected}, a key ingredient for the proof of the main theorem.  In Section \ref{main}, we state the remaining three conditions and prove the main result, Theorem \ref{mainthm}, that a marginally trapped tube must form, have $i^+$ as a limit point, and be achronal and connected near $i^+$.  
Theorem \ref{pricelaw} shows how Price law decay implies that the marginally trapped tube has finite length.  
In Section \ref{aps}, we give some applications of Theorem \ref{mainthm}:  
first we discuss how Theorems \ref{mainthm} and \ref{pricelaw} apply to the (ingoing) Vaidya spacetime, then we turn to self-gravitating Higgs field spacetimes, presenting two different ways in which Theorem \ref{mainthm} can be applied.  
In Theorem \ref{HiggsR}, we postulate an explicit inverse power decay rate in order to extract the hypotheses of Theorem \ref{mainthm}, whereas in Theorem \ref{HiggsM} we make only certain smallness and monotonicity assumptions (which are nonetheless quite strong).

\medskip

\noindent
\bf
Acknowledgment:
\rm
The author thanks Mihalis Dafermos for proposing the problem and making many illuminating suggestions; Lars Andersson for bringing to our attention the question of the affine length of the MTT; the Isaac Newton Institute in Cambridge, England for providing an excellent research environment during the programs on Global Problems in Mathematical Relativity; and an anonymous referee for a number of helpful comments. This paper constitutes a portion of the author's doctoral thesis under advisor Dan Pollack, for whose assistance and support she is enormously grateful.

\section{Background assumptions}\label{ba}
\subsection{Spherical symmetry \& the initial value problem}\label{ssivp}
\medskip
\noindent

\medskip
\indent
A spacetime $(M,g)$ is said to be spherically symmetric if the Lie group $SO(3)$ acts on it by isometries with orbits which are either fixed points or spacelike 2-spheres.  If we assume that the quotient $\mathcal{Q} = M/SO(3)$ is a manifold with (possibly empty) boundary $\Gamma$  corresponding to the points fixed by the $SO(3)$-action,  that $\mathcal{Q}$  inherits a 1+1-dimensional Lorentzian structure, and that its  topology  is such that we may conformally embed it into Minkowski space $(\mathbb{R}^2, \eta)$, then the image of this conformal embedding retains the important causal and asymptotic features of the original spacetime.  In particular, such features of $(M,g)$ as black holes, event horizons, and marginally trapped tubes are preserved and may be studied at this quotient level.

Specifically, we fix double null coordinates $(u,v)$ on $\mathbb{R}^2$, such that the Minkowski metric $\eta$ takes the form $\eta = -du \, dv$ and such that $(\mathbb{R}^2, \eta)$ is time oriented in the usual way, with $u$ and $v$ both increasing toward the future. (In our Penrose diagrams, we will always depict the positive $u$- and $v$-axes at $135^\circ$ and $45^\circ$ from the usual positive $x$-axis, respectively.)  With respect to a conformal embedding, the metric on a 1+1 Lorentzian quotient manifold $\mathcal{Q}$ as above then takes the form $-\Omega^2 du \, dv$, where $\Omega=\Omega(u,v)$ is a smooth positive function on $\mathcal{Q}$.  Suppressing pullback notation, the original metric $g$ may be expressed 
\begin{equation} g = -\Omega^2 du \, dv + r^2 g_{S^2}, \label{sphg} \end{equation} 
where  $g_{S^2} = d\theta^2 + \sin^2\!\theta \,d\varphi^2$ is the standard metric on $S^2$,  and the radial function $r$ is a smooth nonnegative function on $\mathcal{Q}$ such that $r(q) = 0$ if and only if $q \in \Gamma$.  
Choosing units so that the Einstein field equations take the form
\[ R_{\alpha \beta} - \textstyle\frac{1}{2}Rg_{\alpha \beta} = 2T_{\alpha \beta}, \]
by direct computation we find that the field equations for the metric (\ref{sphg}) on $M$ yield the following system of pointwise equations on $\mathcal{Q}$:
\begin{eqnarray}  
\del_u(\Omega^{-2} \del_u r) & = & - r \Omega^{-2} T_{uu} \label{eq1}\\
\del_v(\Omega^{-2} \del_v r) & = & - r \Omega^{-2} T_{vv} \label{eq2}\\
\del_u m & = & 2 r^2 \Omega^{-2} (T_{uv} \del_u r - T_{uu} \del_v r) \label{eq3}\\
\del_v m & = & 2 r^2 \Omega^{-2} (T_{uv} \del_v r - T_{vv} \del_u r) \label{eq4},  
\end{eqnarray}
where $T_{uu}$, $T_{uv}$, and $T_{vv}$ are component functions of the stress-energy tensor $T_{\alpha \beta}$ on $M$ and 
\begin{equation} m = m(u,v) = \frac{r}{2}(1 + 4\Omega^{-2} \del_u r \, \del_v r) \label{m}\end{equation} 
is the Hawking mass.  Note that the null constraints (\ref{eq1}) and (\ref{eq2}) are just Raychaudhuri's equation applied to each of the two null directions in $\mathcal{Q}$.

The study of spherically symmetric 3+1-dimensional spacetimes is thus essentially equivalent to the study of conformal metrics paired with radial functions on subsets of $(\mathbb{R}^2, \eta)$, and the relative simplicity of the latter recommends it as a starting point.  Without \it a priori \rm knowledge about an ``upstairs" spacetime $(M,g)$ or the embedding $\mathcal{Q} \hookrightarrow \mathbb{R}^2$, it is natural to begin with a generalized initial value problem for the system (\ref{eq1})-(\ref{eq4}).  

\begin{figure}[hbtp]
\begin{center}
{
\psfrag{G}{$\mathcal{G}(u_0,v_0)$}
\psfrag{K}{$K(u_0,v_0)$}
\psfrag{Cout}{$\mathcal{C}_{out}$}
\psfrag{Cin}{$\mathcal{C}_{in}$}
\psfrag{i+}{$i^+$}
\resizebox{1.81in}{!}{\includegraphics{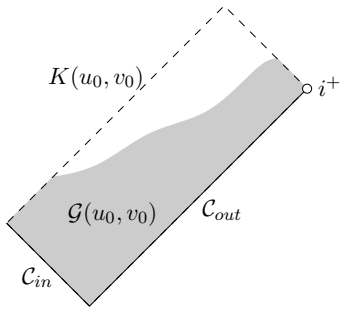}}
}
\caption{\emph{The characteristic rectangle $K(u_0, v_0)$ contains the spacetime $\mathcal{G}(u_0, v_0)$.  In applications to specific matter models, $\mathcal{G}(u_0, v_0)$ would be the maximal future development of initial data prescribed on $\mathcal{C}_{in}\cup\mathcal{C}_{out}$.}}
\label{K}
\end{center}
\end{figure}

First, for any  values $u, v > 0$, we use $K(u,v)$ to denote the characteristic rectangle given by
\[ K(u,v) = [0, u] \times [v, \infty). \]
Next, suppose we choose some values $u_0, v_0 > 0$, \it fix \rm the specific rectangle \linebreak $K(u_0, v_0)$, and define initial hypersurfaces $\mathcal{C}_{in} = [0, u_0] \times \{ v_0 \}$ and $\mathcal{C}_{out} = \{ u_0 \} \times [v_0, \infty)$.  In a specific matter model, we would prescribe initial data for the metric and the stress-energy tensor along $\mathcal{C}_{in}$ and $\mathcal{C}_{out}$ in such a way that the four equations (\ref{eq1})-(\ref{eq4}) were satisfied, then use these equations coupled to the evolution equation(s) associated with the matter model to establish the existence of a maximal future development of the initial data.  In our situation, however, we are working with a general stress-energy tensor with no evolution equations of its own beyond those imposed by the Bianchi identity, $\text{div}_gT = 0$, so we will instead assume only that we have a globally hyperbolic (relatively) open spacetime $\mathcal{G}(u_0, v_0) \subset K(u_0, v_0)$ such that $\mathcal{C}_{in} \cup \mathcal{C}_{out} \subset \mathcal{G}(u_0, v_0)$.  For simplicity we also assume that our rectangle does not intersect the center of symmetry along its two past edges, i.e.\ $r > 0$ on $\mathcal{C}_{in} \cup \mathcal{C}_{out}$; in practice one could always shrink the rectangle so as to insure this.  See Figure \ref{K} for a representative Penrose diagram.

\subsection{Black hole \& energy assumptions}\label{bhea}
\medskip
\noindent

\medskip
\indent
In this section, we make a number of assumptions to which we will refer later, namely in the statement and proof of our main result.  These assumptions are the basic requirements that the spacetime and stress-energy tensor must satisfy in order to be physically reasonable and relevant to black hole spacetimes.   We label them here with upper-case Roman numerals for convenience.  

\medskip

First, on physical grounds, we want the ``upstairs" stress-energy tensor $T_{\alpha \beta}$ to satisfy the dominant energy condition.  In the 1+1-setting, this condition yields the following pointwise inequalities at the quotient level:
\bigskip

\noindent {\bf{\DEC}} \hspace*{1.2in}  $T_{uu} \geq 0, \,\,\,\, T_{uv} \geq 0, \hspace*{.1in}\text{and}\hspace*{.1in} T_{vv} \geq 0.$

\bigskip

Second, because we are not working with a specific global existence result in an evolutionary setting, we explicitly require that the spacetime $\mathcal{G}(u_0, v_0)$ obtained in the previous section be a past subset of $K(u_0, v_0)$, i.e.
\bigskip

\noindent {\bf{\past}} \hspace*{1.4in}  $J^-(\mathcal{G}(u_0, v_0)) \subset \,\mathcal{G}(u_0, v_0)$.  

\bigskip

Next, we assume that along $\mathcal{C}_{out}$ the functions $r$ and $m$ satisfy

\bigskip

\noindent {\bf{\rbounds}} \hspace*{1.95in}  $r \leq r_+$, 

\bigskip

and

\bigskip

\noindent {\bf{\mpos}} \hspace*{1.75in}  $0 \leq m \leq m_+$,

\bigskip

\noindent
where the constants $r_+$, $m_+ < \infty$ are chosen to be the respective suprema of $r$ and $m$ along $\mathcal{C}_{out}$.  
Since our aim is to say something about the interiors of spherically symmetric black holes, we want to choose the data on the initial hypersurface $\mathcal{C}_{out}$ of our characteristic rectangle $K(u_0, v_0)$ to insure that they would agree with that we would find along (the quotient of) the event horizon in a general spherically symmetric black hole spacetime --- assumption III  provides that correspondence.  
In particular,  the boundedness of $r$ along $C_{out}$ is precisely the requirement that $\mathcal{C}_{out}$ must satisfy in order to lie inside a black hole in (the quotient of) an asymptotically flat spacetime, at least provided the black hole has bounded surface area (or equivalently, bounded `entropy').  
Indeed, in the general spherically symmetric setting described in \cite{D2}, whether $r$ is bounded or tends to infinity along outgoing null rays is precisely what distinguishes the black hole region from the domain of outer communications, respectively.
Since $\mathcal{C}_{out}$ is necessarily defined for arbitrarily large values of $v$, while outgoing null rays past the event horizon need not be, it is natural to interpret $\mathcal{C}_{out}$ as lying along the event horizon itself.  
For this reason we will often refer to its terminal point $(0, \infty)$ --- which is not in the spacetime, strictly speaking --- as $i^+$, future timelike infinity.  The first inequality of \mpos\, is physically natural, since it just requires that the quasi-local mass $m$ be nonnegative, while the second inequality is actually slightly redundant: given equation (\ref{m}), the boundedness of $m$ along $\mathcal{C}_{out}$ follows immediately from the fact that it is nonnegative and that $r$ is bounded. Indeed, we must have the relation $m_+ \leq \frac{1}{2}r_+$.

\medskip

For the next two of our assumptions, we must introduce notions of trappedness in our spacetime.  Each point $(u,v)$ of $\mathcal{G}(u_0, v_0)$ represents a 2-sphere of radius $r=r(u,v)$ in the original manifold $M$, and the two future null directions orthogonal to this sphere are precisely $\del_u$ and $\del_v$.  We designate $u$ as the ``ingoing" direction and $v$ the ``outgoing" direction and use $\theta_-$ and $\theta_+$ to denote the expansions in the directions $\del_u$ and $\del_v$, respectively.  The induced Riemannian metric on this two-sphere is of course just $h_{ab} = r^2 (g_{S^2})_{ab}$, and a straightforward calculation  shows that $\theta_- = 2 (\del_u r) r^{-1}$ and $\theta_+ = 2 (\del_v r)r^{-1}$.  Since $r$ is strictly positive away from the center of symmetry $\Gamma$, the signs of $\theta_+$ and $\theta_-$ are exactly those of $\del_v r$ and $\del_u r$, respectively.  This fact motivates the definitions of the following three subsets of interest: the \textit{regular region} 
\[ \mathcal{R} = \{ (u,v) \in \mathcal{G}(u_0, v_0) : \del_v r > 0 \text{ and } \del_u r < 0 \}, \] the \textit{trapped region} 
\[ \mathcal{T} = \{ (u,v) \in \mathcal{G}(u_0, v_0) : \del_v r < 0 \text{ and } \del_u r < 0 \}, \] and the \textit{marginally trapped tube}, 
\[ \mathcal{A} = \{ (u,v) \in \mathcal{G}(u_0, v_0) : \del_v r = 0 \text{ and } \del_u r < 0 \}. \] Note that $\mathcal{A}$ is in fact a hypersurface of 
$\mathcal{G}(u_0, v_0)$ provided that $0$ is a regular value of $\del_v r$.  It is then called a tube because ``upstairs" it is foliated by 2-spheres.

For us, an \textit{anti-trapped surface} is one for which $\del_u r \geq 0$.  (Here we are adding the marginal case $\del_u r = 0$ into the usual definition in order to avoid introducing unnecessary terminology.)  We impose the assumption that no anti-trapped surfaces are present initially: 
\medskip

\noindent{\bf{\drduneg}}\hspace*{1.7in} $\del_u r < 0 \text{ along } \mathcal{C}_{out}$.

\medskip

\begin{figure}[hbtp]
\begin{center}
{
\psfrag{Cout}{$\mathcal{C}_{out}$}
\psfrag{Cin}{$\mathcal{C}_{in}$}
\psfrag{i+}{$i^+$}
\psfrag{A}{$\mathcal{A}$}
\psfrag{R}{$\mathcal{R}$}
\psfrag{T}{$\mathcal{T}$}
\resizebox{1.81in}{!}{\includegraphics{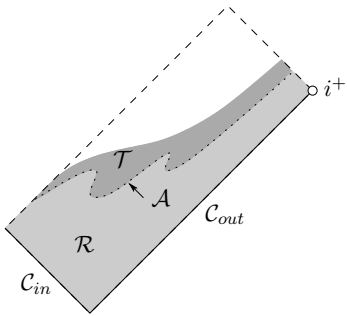}}
}
\caption{\emph{The spacetime $\mathcal{G}(u_0, v_0)$ comprises the regular and trapped regions, $\mathcal{R}$ and $\mathcal{T}$, and a marginally trapped tube $\mathcal{A}$, shown here as a dotted curve.  Note that the marginally trapped tube shown is not achronal, but it does comply with Proposition \ref{handy}.}}
\label{RAT}
\end{center}
\end{figure}

\noindent
 This assumption is motivated primarily by the following result due to Christodoulou (see \cite{C,D2}):
\begin{prop}\label{1} If $\del_u r < 0$ along $\mathcal{C}_{out}$, then $\mathcal{G}(u_0, v_0) = \mathcal{R} \,\cup\, \mathcal{T} \,\cup\, \mathcal{A}$ --- that is, anti-trapped surfaces cannot evolve if none are present initially. 
\end{prop}
\noindent
 See Figure \ref{RAT} for a representative Penrose diagram. 
\begin{proof}
Let $(u, v)$ be any point in $\mathcal{G}(u_0, v_0)$.  Then integrating Raychaudhuri's equation (\ref{eq1}) along the ingoing null ray to the past of  $(u, v)$, we obtain
\[ (\Omega^{-2} \del_u r)(u, v) = (\Omega^{-2} \del_u r)(0, v) -  \int_{0}^{u} r \,\Omega^{-2} T_{uu} (U, v) \, dU. \]
Since we have assumed that $\del_u r < 0 $ along $\mathcal{C}_{out}$ and that $T_{uu} \geq 0$ by assumption \DEC, the righthand side of this equation is strictly negative, and hence so is the lefthand side.
\end{proof}

\noindent
Thus assumption \drduneg\, guarantees that a spacelike marginally trapped tube in \linebreak $\mathcal{G}(u_0, v_0)$  is indeed a dynamical horizon as defined by Ashtekar and Krishnan   \cite{A}, since their definition requires both that $\theta_+ = 0$ and $\theta_- < 0$ along $\mathcal{A}$.

\medskip
One further consequence of the dominant energy condition (assumption \DEC), the Einstein equations (\ref{eq1})-(\ref{eq4}), and our definitions of $\mathcal{A}$ and $\mathcal{T}$ is the following proposition due to Christodoulou, which will be of considerable use later on:

\begin{prop}\label{handy}  If $(u,v) \in \mathcal{T} \cup \mathcal{A}$, then $(u, v^*) \in \mathcal{T} \cup \mathcal{A}$  for all $v^* > v$.  Similarly, if
$(u,v) \in \mathcal{T}$, then $(u, v^*) \in \mathcal{T}$  for all $v^* > v$.
\end{prop}

\begin{proof} Integrating Raychaudhuri's equation (\ref{eq2}) along the null ray to the future of $(u,v)$ yields
\[ (\Omega^{-2} \del_v r)(u, v^*) = (\Omega^{-2} \del_v r)(u, v) - \int_{v}^{v^*} r \,\Omega^{-2} T_{vv} (u, V) \,dV \]
for $v^* > v$.  Since $T_{vv} \geq 0$ everywhere by assumption \DEC, the righthand side of this equation will be nonpositive if $(\Omega^{-2} \del_v r)(u, v) \leq 0$, and strictly negative if
$(\Omega^{-2} \del_v r)(u, v) < 0$.  Since $\Omega > 0$ everywhere, both statements of the proposition now follow immediately. 
\end{proof}

In a black hole spacetime, the trapped region $\mathcal{T}$ must be contained inside the black hole.  Since we would like $\mathcal{C}_{out}$ to represent an event horizon, we must therefore require that $\del_v r \geq 0$ along $\mathcal{C}_{out}$.  Combining this inequality with Proposition \ref{handy}, we see that if $\mathcal{A}$ intersects $\mathcal{C}_{out}$ at a single point, then the two must in fact coincide to the future of that point.  This is indeed the case in the Schwarzschild and Reisner-Nordstr\"{o}m spacetimes, in which the black hole coincides exactly with the trapped region $\mathcal{T}$.  However, we are really only interested in the cases in which the marginally trapped tube $\mathcal{A}$ does \it not \rm coincide with the event horizon, so we will instead assume

\medskip

\noindent{\bf{\drdvpos}} \hspace*{1.57in} $0 < \del_v r$ along $\mathcal{C}_{out}$.

\medskip

\noindent
It is now clear that the values $r_+$ and $m_+$ specified in assumptions \rbounds\, and \mpos\, are in fact the asymptotic values of $r$ and $m$ along $\mathcal{C}_{out}$, respectively (the monotonicity of $m$ follows from assumption \DEC\, and equation (\ref{eq4})).

\medskip

Finally, we will need to assume that $\mathcal{G}(u_0, v_0)$ satisfies the extension principle formulated in \cite{D2}.  This principle holds for self-gravitating Higgs fields and self-gravitating collisionless matter \cite{D3, DR2}, and it is expected to hold for a number of other physically reasonable models \cite{D2}.  Regarding set closures as being taken with respect to the topology of $K(u_0,v_0)$, the extension principle may be formulated as follows:

\bigskip

\noindent {\bf{\extprinc}} \,\,  If $p \in \overline{\mathcal{R}},$ and $q \in \overline{\mathcal{R}}\cap I^-(p)$ such that $J^-(p) \cap J^+(q)\!\setminus\!\!\{p\} \subset \mathcal{R} \cup \mathcal{A}$, 

\begin{figure}[hbtp]
\begin{center}
{
\psfrag{q}{$q$}
\psfrag{p}{$p$}
\psfrag{qu}{?}
\psfrag{RuA}{$\mathcal{R}\cup\mathcal{A}$}
\resizebox{.9in}{!}{\includegraphics{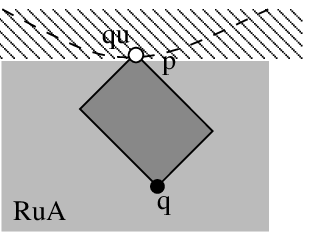}}
}
\end{center}
\end{figure}

\hspace*{.42in} then $p \in \mathcal{R} \cup \mathcal{A}$.    

\begin{figure}[hbtp]
\begin{center}
{
\psfrag{q}{$q$}
\psfrag{p}{$p$}
\psfrag{RuA}{$\mathcal{R}\cup\mathcal{A}$}
\resizebox{.9in}{!}{\includegraphics{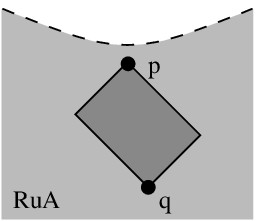}}
}
\end{center}
\end{figure}

\noindent 
Note that the formulation of the extension principle given in  \cite{D2} specifically excludes the possibility that the point $p \in \overline{\Gamma}$, where $\Gamma$ denotes the center of symmetry of $\mathcal{G}(u_0, v_0)$, i.e.\ the set of points at which $r=0$.  However, in our situation, Proposition \ref{handy} and our assumption that $r > 0$ on $\mathcal{C}_{in} \cup \mathcal{C}_{out}$ together imply that $r > 0$ everywhere in $\overline{R}$, so our formulation is slightly simpler.

\section{Achronality \& connectedness}\label{aandc}

The proof of our main result, Theorem \ref{mainthm}, which appears in Section \ref{asym}, relies on two more general propositions, both of which are of interest in their own right.  These propositions require a weak version of one of the four conditions appearing in Theorem \ref{mainthm}, a condition which we now state separately:

\bigskip

\noindent{\bf{\Acond}}\hspace*{1.9in} $T_{uv} \Omega^{-2} < \displaystyle\frac{1}{4r^2}$.

\medskip
\noindent
In practice, we will only require that condition \Acond\, hold in some small subset of our spacetime $\mathcal{G}(u_0, v_0)$. 
The expression $T_{uv}\,\Omega^{-2}$ takes a particularly simple form in many matter models.  For a perfect fluid of pressure $P$ and energy density $\rho$, it is the quantity $\frac{1}{4}(\rho - P)$. For a self-gravitating Higgs field $\phi$ with potential $V(\phi)$, it is  $\frac{1}{2}V(\phi)$. And for an Einstein-Maxwell massless scalar field of charge $e$, it is $\frac{1}{4}e^2 r^{-4}$. 

%%%%%%%%%%%%%%%%%%%%%%%%%%%%%% ACHRONAL %%%%%%%%%%%%%%%%%%%%%%%%%%%%%%%%%%%%%

\begin{prop}\label{achronal} Suppose $(\mathcal{G}(u_0, v_0), -\Omega^2\,du\,dv)$ is a spacetime obtained as in Section \ref{ssivp} with radial function $r$, and suppose it satisfies assumptions \rm \DEC-\extprinc\, \it  of Section \ref{bhea}.  If $\mathcal{A}$ is nonempty and condition \rm \Acond\, \it holds in $\mathcal{A}$, then each of its connected components is achronal with no ingoing null segments.
\end{prop}

\noindent \it Remark: \rm
In \cite{B}, Booth et al. give a necessary and sufficient condition for a general marginally trapped tube (not necessarily spherically symmetric) to be achronal; that condition is precisely \Acond\, in our setting.  The proof of Proposition \ref{achronal} essentially duplicates their reasoning, although it is formulated somewhat differently.

\medskip
\begin{proof}  To begin, we must establish that $\mathcal{A}$ is in fact a hypersurface in $\mathcal{G}(u_0, v_0)$. Since $\mathcal{A}$ is defined as a level set, this is equivalent to showing that $0$ is a regular value of $\del_v r$, i.e. that the differential $D(\del_v r)$ is non-degenerate at points where $\del_v r = 0$.  Since $D(\del_v r)$ has components $\del^{2}_{uv}r$ and $\del^2_{vv}r$, it suffices to show that $\del^{2}_{uv}r < 0$ along $\mathcal{A}$.  

Rearranging equation (\ref{eq3}) and then combining with equations (\ref{eq1}) and (\ref{m}) yields
\begin{eqnarray*} 2 r^2 \Omega^{-2} T_{uv} \del_u r & = & 2 r^2 \Omega^{-2} T_{uu} \del_v r + \del_u m \\ 
& = & \textstyle\frac{1}{2}( \del_u r) + 2(\del_u r )^2( \del_v r)\Omega^{-2} + 2 r (\del^{2}_{uv} r)\Omega^{-2}\, \del_u r, 
\end{eqnarray*}
and solving for $\del^{2}_{uv} r$, we obtain
\begin{equation}
\del^{2}_{uv} r =  -\textstyle\frac{1}{2}\Omega^2 r^{-2} \alpha, \label{druv}
\end{equation}
where $\alpha$ is given by
\begin{equation} \alpha = m - 2r^3\,\Omega^{-2}T_{uv}.\label{alpha}\end{equation}
Since $r$ and $\Omega$ are strictly positive in $\mathcal{G}(u_0, v_0)$, it is enough to show that $\alpha > 0$ along $\mathcal{A}$.
Using condition \Acond \, and the fact that $\del_v r = 0$ on $\mathcal{A}$, we have
\begin{eqnarray*} \alpha & = & m - 2r^3\,\Omega^{-2}T_{uv} \\
& = & \textstyle\frac{r}{2} - 2r^3\,\Omega^{-2}T_{uv} \\
& = & \textstyle\frac{r^3}{2}(\frac{1}{r^2} - 4\,\Omega^{-2}T_{uv}) \\
& > & 0,
\end{eqnarray*} 
so  $\del^{2}_{uv} r < 0$ along $\mathcal{A}$ as desired.

Now, we have $\mathcal{A} \neq \emptyset$, and since we now know it is a 1-dimensional submanifold of the spacetime,  we can parameterize any connected component of it by a curve $\gamma(t)=(u(t), v(t))$.  Since $\del_v r \equiv 0$ along $\mathcal{A}$, at points on $\mathcal{A}$ we have
\begin{equation} \dot{\gamma} (\del_v r) = 0 = \frac{du}{dt} (\del^{\,2}_{uv} r) + \frac{dv}{dt} (\del^2_{vv} r). \label{dvdu}\end{equation}
By the result of Proposition \ref{handy}, we know that we can describe all but the outgoing null segments of $\mathcal{A}$ in terms of a function $v(u)$, defined on some (possibly disconnected) subset of $[0, u_0]$.  From equation (\ref{dvdu}) we see that its slope is given by
\begin{equation} \frac{dv}{du} = \frac{dv/dt}{du/dt} = - \frac{\del^{\,2}_{uv} r}{\del^{\,2}_{vv} r} \label{slope}\end{equation}
at points where $\del^2_{vv} r \neq 0$.  

Showing that $\mathcal{A}$ is achronal thus amounts to showing that this slope $\frac{dv}{du} \leq 0$ wherever it is defined, since  the points at which the slope is not defined correspond to points on outgoing null segments.  In fact, we will show that, where it is defined, $\frac{dv}{du} < 0$, thereby excluding the possibility of ingoing null segments.  

Expanding the lefthand side of (\ref{eq2}), we see that along $\mathcal{A}$, 
\[ \Omega^{-2} \del^{\,2}_{vv} r = -r\,\Omega^{-2}T_{vv},\]
or rather,
\begin{equation} \del^{\,2}_{vv} r = -r T_{vv}. \label{drvv} \end{equation}  Substituting equations (\ref{drvv}) and (\ref{druv}) into equation (\ref{slope}) then yields
\begin{eqnarray*} \frac{dv}{du} & =  & -\frac{\Omega^2 \alpha}{2r^3 T_{vv}}.
\end{eqnarray*}
Since $r$, $\Omega$, and $\alpha$ are all positive along $\mathcal{A}$ and $T_{vv}$ is nonnegative by assumption  \DEC, the dominant energy condition, we conclude that $\frac{dv}{du} < 0$ at points along $\mathcal{A}$ at which $T_{vv} > 0$, which is exactly what was needed. 
\end{proof}

%%%%%%%%%%%%%%%%%%%%%%%%%%%%%%%%%%%%%%%%%%%%%%%%%% CONNECTED %%%%%%%%%%%%%%%%%%%%%%%%%%%%%%%%%%%%%%%%%%%%%%%%

\bigskip

For technical reasons, in the next proposition we use $\overline{K}(u_0, v_0)$ to denote the ``compactification"  of our initial rectangle $K(u_0, v_0)$, that is, $\overline{K}(u_0, v_0) = [0, u_0] \times [v_0, \infty]$.  Set closures are taken with respect to $\overline{K}$ rather than $K$ so as to include points at infinity (the Cauchy horizon).  

We also want to confine ourselves to regions of the spacetime in which $r$ is close to $r_+$, so for any $\delta > 0$, let \[ \mathcal{W}(\delta) = \{ (u,v) \in \mathcal{G}(u_0, v_0) : r(u,v) \geq r_+ - \delta \}. \]  

\begin{prop}\label{connected} Suppose $(\mathcal{G}(u_0, v_0), -\Omega^2\,du\,dv)$ is a spacetime obtained as in Section \ref{ssivp} with radial function $r$, suppose it satisfies  assumptions \rm \DEC-\extprinc\, \it of Section \ref{bhea}, and suppose there exists $\delta > 0$ such that for $\mathcal{W} = \mathcal{W}(\delta)$,  condition \rm \Acond\, \it is satisfied in $\mathcal{A} \cap \mathcal{W}$.
If $\mathcal{G}(u_0, v_0)$ does not contain a marginally trapped tube which is asymptotic to the event horizon, then  $\mathcal{W} \cap \mathcal{R}$ contains a rectangle $K(u_1,v_1)$ for some $u_1 \in (0,u_0]$, $v_1 \in [v_0, \infty)$.
\end{prop}

\smallskip
\noindent{\it{Proof.}}\, The proposition is an immediate consequence of the following two lemmas: 

\smallskip
\begin{lem}\label{lem1} Under the hypotheses of Proposition \ref{connected}, if $\mathcal{A} \cap \mathcal{W} \neq \emptyset$, then $\mathcal{A} \cap \mathcal{W}$ is connected and terminates either at $i^+$ (in which case it is asymptotic to the event horizon) or along the Cauchy horizon $(0, u_0] \times \{ \infty \} \subset \overline{K}(u_0, v_0)$.  In the latter case, $\mathcal{W} \cap \mathcal{R}$ contains a rectangle $K(u_1,v_1)$ for some $u_1 \in (0,u_0]$, $v_1 \in [v_0, \infty)$.
\end{lem}

\begin{lem}\label{lem2} Under the hypotheses of Proposition \ref{connected}, if $\mathcal{A} \cap \mathcal{W} = \emptyset$, then $\mathcal{W} \cap \mathcal{R}$ contains a rectangle $K(u_1,v_1)$ for some $u_1 \in (0,u_0]$, $v_1 \in [v_0, \infty)$.
\end{lem}

\it \noindent Remark. \rm The existence of the rectangle in $\mathcal{W} \cap \mathcal{R}$ is what is required for the proof of the main result, but the first statement of Lemma \ref{lem1} establishing the connectedness of $\mathcal{A}$ is of independent interest.  As we will see in the proof of the lemma, that result hinges on the extension principle (\extprinc) and the achronality of $\mathcal{A}$.   

%%%%%%%%%%%%%%%%%%%%%%%%%%%%%%%%%%%%%%%%%%%%% LEMMA 1 %%%%%%%%%%%%%%%%%%%%%%%%%%%%%%%%%%%%%%%%%%%%%%%%%%%

\medskip
\noindent
\begin{proof}[\it{Proof of Lemma \ref{lem1}}] First we lay some groundwork.  Let $\mathcal{S}$ denote any connected component of $\del\mathcal{W} \cap \mathcal{G}(u_0, v_0)$, that is, a connected component of the level set $\{ r = r_+ - \delta \}$. Since  $\del_u r$ is strictly negative by Proposition \ref{1}, the differential $Dr$ is nondegenerate in all of $\mathcal{G}(u_0, v_0)$, and thus $\mathcal{S}$ is a smooth curve segment whose endpoints lie on $\del\mathcal{G}(u_0, v_0)$.  Parameterizing $\mathcal{S}$ by a curve $\gamma(t) = (u(t), v(t))$, we compute that
\begin{equation} 0 = \dot{\gamma}(r) = (\del_u r)\textstyle\frac{du}{dt} + (\del_v r)\frac{dv}{dt}. \label{rtan}\end{equation}
Now, $\del_v r > 0$ in $\mathcal{R}$ and $\del_u r< 0$ everywhere, so $\frac{du}{dt}$ and $\frac{dv}{dt}$ must have the same sign in $\mathcal{R}$, which in turn implies that 
\[ \langle \dot{\gamma}, \dot{\gamma} \,\rangle = -\Omega^2 \,du\,dv \,(\dot{\gamma}, \dot{\gamma}) = -\Omega^2 (\textstyle\frac{du}{dt})(\frac{dv}{dt}) < 0\]
--- that is, $\mathcal{S} \cap \mathcal{R}$ is timelike.  Furthermore, equation (\ref{rtan}) implies that if $\mathcal{S}$ intersects $\mathcal{A}$ and passes into $\mathcal{T}$, it is $\frac{du}{dt}$ that changes sign, while $\frac{dv}{dt}$ does not.  Hence $\mathcal{S} \cap (\mathcal{R} \cup \mathcal{A})$ must be causal with no ingoing null segments.

Now let $\mathcal{S}_{0}$ denote the connected component of $\del\mathcal{W} \cap \mathcal{G}(u_0, v_0)$
that intersects one of the initial hypersurfaces $\mathcal{C}_{in}$ or $\mathcal{C}_{out}$.  (If no component intersects $\mathcal{C}_{in} \cup \mathcal{C}_{out}$, then without loss of generality, we may shrink $\delta$ until one does.  And by the monotonicity of $r$ along each initial hypersurface, we can be sure that there is at \it most \rm one such component.)  The above characterization of the causal behavior of $\del \mathcal{W}$ implies that all of $\mathcal{S}_0$ lies to the future of this endpoint on $\mathcal{C}_{in}$ or $\mathcal{C}_{out}$. 
This past endpoint lies in $\mathcal{R}$ since both initial hypersurfaces do, so $\mathcal{S}_0 \cap (\mathcal{R} \cup \mathcal{A})$ is nonempty and, by the preceding argument, causal.  Let $q$ denote its future endpoint in $\overline{\mathcal{G}(u_0, v_0)}$.  
Then $J^+(q)\setminus\{q\} \cap \mathcal{W} = \emptyset$.  To see this, observe that if 
$q \in \overline{\mathcal{G}(u_0, v_0)}\setminus \mathcal{G}{(u_0, v_0)}$, then the fact that $\mathcal{G}(u_0, v_0)$ is a past set (assumption \past) in fact implies that $J^+(q) \cap \mathcal{G}(u_0, v_0) = \emptyset$.  Otherwise, 
$q \in \mathcal{G}(u_0, v_0)$ and hence $q \in \mathcal{A} \cap \mathcal{S}_0$, so by the choice of $q$ and Proposition \ref{handy}, the outgoing null ray to the future of $q$ must lie in the trapped region, i.e. 
\[ \{ u(q) \} \times (v(q), \infty) \cap \mathcal{G}(u_0, v_0) \subset \mathcal{T}.\] 
Since $r(q) = r_+ - \delta$ and $r$ is strictly decreasing along  
$\{ u(q) \} \times [v(q), \infty) \cap \mathcal{G}(u_0, v_0)$,  
\[ \{ u(q) \} \times (v(q), \infty) \cap \mathcal{G}(u_0, v_0) \cap \mathcal{W} = \emptyset. \]
Then the inequality $\del_u r < 0$ guarantees that $J^+(q)\setminus\{q\} \cap \mathcal{W} = \emptyset$.

Let $\mathcal{A}_0$ denote any connected component of $\mathcal{A} \cap \mathcal{W}$. We will show that it either terminates at $i^+$ or along the Cauchy horizon, after which the connectedness of $\mathcal{A} \cap \mathcal{W}$ and the statement of the lemma will follow.  Since $\mathcal{A}_0$ is a connected curve segment, it has endpoints $p_0$ and $p_1$ in $\overline{\mathcal{G}(u_0, v_0)}$. By Proposition \ref{achronal}, $\mathcal{A}_0$ is achronal with  no ingoing null segments, so if $p_0 \neq p_1$, then one of $p_0$ and $p_1$ must have $v$-coordinate strictly larger than the other --- without loss of generality, suppose it's $p_1=(u_*, v_*)$.  (See Figure \ref{lem1afig} for a Penrose diagram depicting $\mathcal{A}_0$ and $\mathcal{S}_0$.)  Then for any point $(u,v) \in \mathcal{A}_0$, the achronality of $\mathcal{A}_0$ implies that $u \geq u_*$ and $v \leq v_*$.  Our goal is to show that $v_* = \infty$, for then
$\mathcal{A}_0$  terminates either at $i^+ = (0, \infty)$ or along the Cauchy horizon $(0, u_0] \times \{ \infty \}$.

\begin{figure}[hbtp]
\begin{center}
{
\psfrag{p0}{$p_0$}
\psfrag{p1}{$p_1$}
\psfrag{q}{$q$}
\psfrag{W0}{$\mathcal{S}_0$}
\psfrag{A0}{$\mathcal{A}_0$}
\psfrag{W}{$\mathcal{W}$}
\psfrag{?}{?}
\psfrag{Cout}{$\mathcal{C}_{out}$}
\psfrag{Cin}{$\mathcal{C}_{in}$}
\psfrag{i+}{$i^+$}
\psfrag{CH}{Cauchy horizon}
\resizebox{2in}{!}{\includegraphics{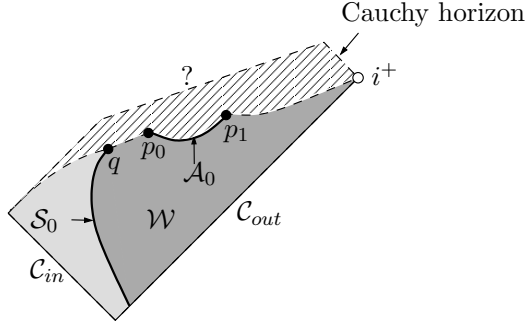}}
}
\caption[\emph{The curves $\mathcal{S}_0$ and $\mathcal{A}_0$ in the proof of Lemma \ref{lem1}}]{A priori, \emph{the curves $\mathcal{S}_0$ and $\mathcal{A}_0$ used in the proof of Lemma \ref{lem1} may be situated as shown.  Their endpoints $q$, $p_0$ and $p_1$ may lie in the spacetime $\mathcal{G}(u_0, v_0)$ or in its boundary, $\overline{\mathcal{G}(u_0, v_0)}\!\!~\setminus~\!\!\mathcal{G}(u_0, v_0)$.  The dashed curves and diagonal lines indicate boundaries and regions which also may or may not be part of the spacetime.  Lemma \ref{lem1} shows that the point $p_1$ must in fact lie along the Cauchy horizon.}}
\label{lem1afig}
\end{center}
\end{figure}

Now, this point $p_1$ is either contained in the spacetime $\mathcal{G}(u_0, v_0)$ itself or in its boundary, $\overline{\mathcal{G}(u_0, v_0)} \setminus \mathcal{G}(u_0, v_0)$, and we must employ different arguments for each case.  In the former case we will deduce that the point $p_1$ coincides with $q$, the future endpoint of the curve $\mathcal{S}_0 \cap (\mathcal{R} \cup \mathcal{A})$, and from there derive a contradiction to how $p_1$ and $q$ were chosen.  In the latter case, we will use the extension principle, assumption \extprinc, to show that if $v_* < \infty$, then $p_1$ is in the spacetime, a contradiction.  Thus we will conclude that $p_1$ must indeed lie on the Cauchy horizon or coincide with $i^+$.

To begin, suppose that $p_1 \in \mathcal{G}(u_0, v_0)$.  Then $p_1 \in \mathcal{A}_0 \cap \del \mathcal{W}$, and in particular, $r(p_1) = r_+ -\delta$.  
Since $\overline{\mathcal{S}_0}$ is causal with no ingoing null segments, the $v$-coordinate of its endpoint $q$ must be greater than that of any other point on
$\mathcal{S}_0$, i.e. $v(q) > v(\tilde{q})$ for any $\tilde{q}\in\mathcal{S}_0$, $\tilde{q} \neq q$.
Now, unless $p_1$ lies on $\overline{\mathcal{S}_0}$, $p_1$ must have $v$-coordinate greater than $v(q)$, i.e. $v_* \geq v(q)$ ---  otherwise, the fact that $r(p_1) = r_+ - \delta$ and $r \equiv r_+ - \delta$ along $\mathcal{S}_0$ would contradict assumptions  \drduneg\, and \drdvpos, the strict monotonicity of $r$ along ingoing null rays and along $\mathcal{C}_{out}$.  Furthermore, $p_1 \in \del\mathcal{W}$ implies that $p_1 \in \mathcal{W}$, which in turn yields that if $v_* \geq v(q)$, then either $p_1 = q$ or $u_* < u(q)$ since $J^+(q)\setminus\{q\} \cap \mathcal{W} = \emptyset$.  Thus the only possibilities remaining are that either $p_1 \in \overline{\mathcal{S}_0}$, or $u_* < u(q)$ and $v_* \geq v(q)$.  

If $u_* < u(q)$ and $v_* \geq v(q)$, then since $p_1 \in \mathcal{A}$, the outgoing null ray to the past of $p_1$ must lie in $\mathcal{R} \cup \mathcal{A}$ by Proposition \ref{handy}, but it must also  contain some point $\tilde{q} \in \text{int}(\mathcal{W})$.  Then $r(\tilde{q}) > r_+ - \delta$, but $r(p_1) = r_+ - \delta$ and $\del_v r \geq 0$ in $\mathcal{R} \cup \mathcal{A}$, a contradiction.   

For the case $p_1 \in \overline{\mathcal{S}_0}$, first note that since $D(\del_v r)$ is nondegenerate at $p_1$ (see the proof of Proposition \ref{achronal}), it must be the case that the curve $\mathcal{A}$  leaves $\mathcal{W}$ at $p_1$ as $v$ increases.  Together with the facts that $\mathcal{A}$ is achronal, $\mathcal{S}_0$ is causal, and $p_1$ was chosen to be the endpoint of $\mathcal{A}_0$ with the largest $v$-coordinate, this implies that $p_1=q$.

We are now in a position to derive the contradiction for this case.  First note that both curves $\mathcal{A}$ and $\del\mathcal{W}$ must extend smoothly through $p_1=q$. As noted in the preceding paragraph, by definition of $\mathcal{A}_0$ and $p_1$, the curve $\mathcal{A}$ leaves $\mathcal{W}$ at $p_1$ as $v$ increases.  On the other hand, by definition of $\mathcal{S}_0$ and the characterization of its causal behavior given above, the curve $\del\mathcal{W}$ must leave $\mathcal{R} \cup \mathcal{A}$ at $q$ as $v$ increases, passing into $\mathcal{T}$; in particular, it must become spacelike for $v > v(q)$.  And since $\del\mathcal{W}$ is leaving $\mathcal{R} \cup \mathcal{A}$, this spacelike curve-continuation past $q = p_1$ must lie to the future of that of $\mathcal{A}$ at least locally.  On the other hand, since $\mathcal{A}$ is leaving $\mathcal{W}$ at $p_1 = q$, its continuation must lie to the future of that of $\del\mathcal{W}$, locally.  But the two cannot coincide past $p_1=q$, by the choices of both $p_1$ and $q$, so we have arrived at a contradiction.

Thus $p_1$ cannot lie in $\mathcal{G}(u_0,v_0)$, so we must have $p_1 \in \overline{\mathcal{G}(u_0, v_0)}\setminus\mathcal{G}(u_0, v_0)$.  If $v_* < \infty$, then consider the ingoing null ray to the past of $p_1$, $[0, u_*] \times \{ v_* \}$.  Since the point $(0, v_*) \in \mathcal{G}(u_0, v_0)$ and $\mathcal{G}(u_0, v_0)$ is open, there must exist some smallest $\tilde{u} \in (0, u_*]$ such that $(\tilde{u}, v_*) \notin \mathcal{G}(u_0, v_0)$.  Since $p_1$ is a limit point of $\mathcal{A}$ but is not in the spacetime, we must have $p_0 \neq p_1$, and so since $\mathcal{A}_0$ is achronal with no ingoing segments, we can parameterize a portion of $\mathcal{A}_0$ in a neighborhood of $p_1$ by $(u(v), v)$, $v \in (v_* - \epsilon, v_*]$, some $\epsilon > 0$.  For each value $v \in (v_* - \epsilon, v_*)$, the ingoing null ray to the past of  the point $(u(v), v) \in \mathcal{A} \cap \mathcal{W}$ must be contained in $\mathcal{W} \cap \mathcal{R}$ --- the ray must lie in $\mathcal{W}$ since $\del_u r < 0$ along it, and it must lie in $\mathcal{R}$  since $\del_u \del_v r < 0$ in $\mathcal{A} \cap \mathcal{W}$.  So in particular, if we choose some $\tilde{v} \in (v_* - \epsilon, v_*)$, then $J^-(\tilde{u}, v_*) \cap J^+(0, \tilde{v}) \setminus \{ (\tilde{u}, v_*) \} \subset \mathcal{R} \cup \mathcal{A}$, and hence by the extension principle, we must have $(\tilde{u}, v_*) \in  \mathcal{R} \cup {A}$ as well, a contradiction.  (See Figure \ref{lem1bfig} for a Penrose diagram of this situation.)  Thus we conclude that $v_* = \infty$, and we are done; this is what we wanted to show.

\begin{figure}[hbtp]
\begin{center}
{
\psfrag{p1}{$p_1 = (u_*, v_*)$}
\psfrag{utv*}{$(\tilde{u}, v_*)$}
\psfrag{A0}{$\mathcal{A}_0$}
\psfrag{u=ut}{$u = \tilde{u}$}
\psfrag{Cout}{$\mathcal{C}_{out}$}
\psfrag{v=vt}{$v = \tilde{v}$}
\psfrag{v=v*-e}{$v=v_*-\epsilon$}
\psfrag{v=v*}{$v=v_*$}
\resizebox{1.85in}{!}{\includegraphics{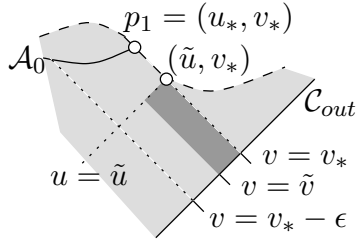}}
}
\caption[\emph{The points $p_1 = (u_*, v_*)$ and $(\tilde{u}, v_*)$ in the proof of Lemma \ref{lem1}}]{\emph{The dashed line indicates the boundary of the spacetime, $\overline{\mathcal{G}(u_0, v_0)}\setminus\mathcal{G}(u_0,v_0)$. The points $p_1 = (u_*, v_*)$ and $(\tilde{u}, v_*)$ lie in this boundary, not in $\mathcal{G}(u_0,v_0)$ itself.  The dark-shaded rectangle is the set $J^-(\tilde{u}, v_*) \cap J^+(0, \tilde{v}) \setminus \{ (\tilde{u}, v_*) \}$ to which we apply the extension principle (assumption {\rm{\extprinc}}) and derive a contradiction.}}
\label{lem1bfig}
\end{center}
\end{figure}

Now we have shown that an arbitrary connected component of $\mathcal{A} \cap \mathcal{W}$ must terminate along the Cauchy horizon $[0, u_0] \times \{ \infty \}$.  To see that this component is unique, i.e.\ $\mathcal{A} \cap \mathcal{W}$ is connected, recall that since  condition \Acond\, holds in $\mathcal{A} \cap \mathcal{W}$, we have $\del_u(\del_v r ) < 0$ there; see the proof of Proposition \ref{achronal}.  This monotonicity in turn implies that any future-directed ingoing null ray intersecting $\mathcal{A} \cap \mathcal{W}$ must pass from $\mathcal{R}$ into $\mathcal{T}$, and so every such ingoing null ray can intersect at most component of $\mathcal{A} \cap \mathcal{W}$ (note that if the ray leaves $\mathcal{W}$, then it cannot reenter it since $\del_u r < 0$).    Therefore, since every connected component of $\mathcal{A} \cap \mathcal{W}$ must exist for arbitrarily large $v$, there can be at most one component.

The last statement of the lemma follows immediately.
\end{proof}

\begin{proof}[\it{Proof of Lemma \ref{lem2}}]
 If $\mathcal{A} \cap \mathcal{W} = \emptyset$, then $\mathcal{W} \subset \mathcal{R}$, so in fact $\mathcal{W} \cap \mathcal{R} = \mathcal{W}$.  Fix a reference point $(u_1, v_1) \in \text{int}(\mathcal{W}) \cap \mathcal{R}$ such that $u_1 > 0$. Note that since $\del_u r < 0$, the past-directed ingoing null ray behind $(u_1, v_1)$ must also be in $\text{int}(\mathcal{W})$, that is, $[0, u_1] \times \{ v_1 \} \subset \text{int}(\mathcal{W})$. 

If $K(u_1, v_1)$ is not wholly contained in $\mathcal{W}$, then there exists some $q \in (\del\mathcal{W}) \cap K(u_1, v_1) \neq \emptyset$, where the boundary $\del \mathcal{W}$ is taken here with respect to $K(u_0, v_0)$ (as opposed to $\mathcal{G}(u_0, v_0)$).  In particular, $q$ cannot lie on $[0, u_1] \times \{ v_1 \}$ by choice of $(u_1, v_1)$, nor can it lie elsewhere in $\mathcal{W} \cap K(u_1, v_1)$, since that would imply that $r(q) = r_+ - \delta$, violating the monotonicity of $r$ in $\mathcal{R}$ ($\del_v r > 0$) and the fact that $r > r_+ - \delta$ on $[0, u_1] \times \{ v_1 \}$.  Hence $q \in \overline{\mathcal{G}(u_0, v_0)}\setminus \mathcal{G}(u_0, v_0)$.  

Fix $v_*$ to be the smallest value in $(v_1, \infty)$ such that $[0, u_1] \times \{ v_* \} \cap (\overline{\mathcal{G}(u_0, v_0)}\setminus\mathcal{G}(u_0, v_0)) \neq \emptyset$, and set $u_*$ to be the smallest value in $(0, u_1]$ such that $(u_*, v_*) \in \overline{\mathcal{G}(u_0, v_0)}\setminus\mathcal{G}(u_0, v_0)$.  Then by construction, the rectangle $J^-(u_*, v_*) \cap J^+(0, v_1) \setminus \{ (u_*, v_*) \} \subset \mathcal{R}$, and so since $r$ is bounded below by $r_+ - \delta$ near $(u_*, v_*)$, the extension principle implies that $(u_*, v_*) \in \mathcal{G}(u_0,v_0)$, a contradiction.

Thus we must in fact have $(\del\mathcal{W}) \cap K(u_1, v_1) = \emptyset$, which implies that $K(u_1, v_1) \subset \mathcal{W} \cap \mathcal{R}$.
\end{proof}

%%%%%%%%%%%%%%%%%%%%%%%%%%%%%%%%%%%%%%%%% MAIN THEOREM %%%%%%%%%%%%%%%%%%%%%%%%%%%%%%%%%%%%%%%%%%%%%%%%%%%%%%%%%%%%%%

\section{Main results}\label{main}

\subsection{Asymptotic behavior}\label{asym}
\medskip
\noindent

\smallskip
We are now ready to state and prove our main result characterizing the asymptotic behavior of certain marginally trapped tubes. 
Afterward we describe how an estimate obtained in the proof of theorem implies that the event horizon of the given spacetime must be future geodesically complete.
We then present a second theorem relating the lengths of such tubes to Price law decay.  

\begin{thm}\label{mainthm}  Suppose $(\mathcal{G}(u_0, v_0), -\Omega^2\,du\,dv)$ is a spacetime obtained as in Section \ref{ssivp} with radial function $r$, and suppose it satisfies the assumptions \rm \DEC-\extprinc\, \it of Section \ref{bhea}.  
Define
\[ \mathcal{W}(\delta) = \{ (u,v) \in \mathcal{G}(u_0, v_0) : r(u,v) \geq r_+ - \delta \} \]
and assume that there exist a constant $0 < c_0 < \frac{1}{4 r_+^2}$, constants $c_1$, $c_2 > 0$, constants $0 < \epsilon < \frac{1}{4r_+^2} - c_0$ and $v^\prime \geq v_0$, and some small $\delta > 0$ such that for $\mathcal{W} = \mathcal{W}(\delta)$ the following conditions hold:
\renewcommand{\arraystretch}{1.5}
\[\begin{array}{cl}
{\bf{\Aprimecond^\prime}}  &  T_{uv}\,\Omega^{-2} \leq c_0 \text{ in } \mathcal{W} \cap \mathcal{R}; \\
{\bf{\Bcondone}} &  T_{uu}/(\del_u r)^2\, \leq c_1 \text{ in } \mathcal{W} \cap \mathcal{R}; \\
{\bf{\Bcondtwo}} & \del_v (\Omega^{-2}T_{uv}) (u, \cdot) \in L^1([v_0, \infty)) \text{ for all } u \in [0, u_0], \text{ and } \\
& \int_{v^\prime}^v \del_v (\Omega^{-2}T_{uv})(u, \tilde{v})\, d\tilde{v} < \epsilon \text{ for all } (u,v) \in \mathcal{W} \cap \mathcal{R} \text{ with } v\geq v^\prime; \\
{\bf{\Ccond}}    &  (-\del_u r)\Omega^{-2} \leq c_2 \text{ along } \mathcal{C}_{out} \cap \mathcal{W}.
\end{array}\]
\noindent
Then the spacetime $\mathcal{G}(u_0, v_0)$ contains a marginally trapped tube $\mathcal{A}$ which is asymptotic to the event horizon, i.e. for every small $u > 0$, there exists some $v > v_0$ such that $(u,v) \in \mathcal{A}$.  Furthermore, for large $v$, $\mathcal{A}$ is connected and achronal with no ingoing null segments. 
\end{thm}

\begin{figure}[hbtp]
\begin{center}
{
\psfrag{A}{\small$\mathcal{A}$}
\psfrag{R}{\small$\mathcal{R}$}
\psfrag{T}{\small$\mathcal{T}$}
\psfrag{Cout}{\small$\mathcal{C}_{out}$}
\psfrag{i+}{\small$i^+$}
\resizebox{1.75in}{!}{\includegraphics{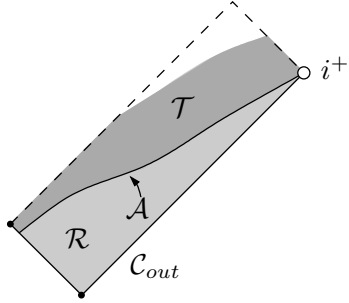}}
}
\caption[\emph{The conclusion of Theorem \ref{mainthm}}]{\emph{Theorem \ref{mainthm} says that the spacetime must contain a (small) characteristic rectangle whose Penrose diagram looks like this -- in particular, the marginally trapped tube $\mathcal{A}$ is achronal and terminates at $i^+$.}}
\label{thm1}
\end{center}
\end{figure}

\noindent
See Figure \ref{thm1} for a representative Penrose diagram.

\noindent \it Remarks: \rm 
The physical meaning of condition \Aprimecond$^\prime$, which is somewhat stronger than condition \Acond, is readily apparent from Proposition \ref{achronal}: it controls the causal behavior of the marginally trapped tube, if one exists.  
Conditions \Bcondone\, and \Bcondtwo\, have no obvious general physical meaning.  
(But note that \Bcondtwo\, is automatically satisfied if $\del_v (\Omega^{-2}T_{uv}) \leq 0$ in $\mathcal{W} \cap \mathcal{R}$.)
Condition \Ccond\, determines a gauge along the event horizon; it may alternately be expressed as saying that the quantities $(1 - \frac{2m}{r})$ and $\del_v r$ approach zero at proportional rates as $v$ tends to infinity along $\mathcal{C}_{out}$, which in turn implies that $m \rightarrow \frac{1}{2}r$ along $\mathcal{C}_{out}$, i.e. that $2m_+ = r_+$.  See also the remarks  concerning conditions \Aprimecond$^\prime$ and \Ccond\, following the proof of the theorem.

These four conditions, as well as the proof of the theorem, were obtained by extrapolating portions of the bootstrap argument in Section 7 of \cite{D1} for Einstein-Maxwell scalar fields.  Conditions \Aprimecond$^\prime$, \Bcondtwo, and \Ccond\, are all satisfied for sufficiently small $\delta$ in \it any \rm  Einstein-Maxwell-scalar field black hole, provided $e < 2m_+$.  (The particular choice of the upper bound for $c_0$ in \Aprimecond$^\prime$ is analogous to the condition in \cite{D1} that the black hole not be ``extremal in the limit,"  i.e. $e < 2m_+$.) Condition \Bcondone\, also holds in all the spacetimes considered in \cite{D1}, but there the proof hinges on the Price law decay imposed on $T_{vv}$; one integrates the scalar field equation by parts and uses the polynomial decay of $T_{vv}$ in the $v$-direction to obtain the bound on $T_{uu}/(\del_u r)^2$. (In fact one obtains something stronger than \Bcondone\, this way, that $T_{uu}/(\del_u r)^2$ decays polynomially with $v$.)  

\begin{proof} 
Our first step is to shrink $\delta$ in order to align with the choice of $\epsilon$.
For this, consider the quantity
\[ \Lambda (\epsilon^*, \delta^*) := \left( \frac{r_+ - \delta^*}{r_+}   \right)^3 \left(\frac{1}{4r_+^2} \left(1- 2\epsilon^* c_2 e^{c_1 r_+ \delta^*}\right) - c_0 \right) - \frac{\delta^*}{r_+}(2c_0 + 3M), 
\]
where
\[ M := \sup_{(u,v)\in\mathcal{G}(u_0,v_0)} \int_{v_0}^v | \del_v (\Omega^{-2}T_{uv}) (u, \tilde{v})| \, d\tilde{v}. \]
Clearly $\Lambda$ is positive  for $\epsilon^*$ and $\delta^*$ sufficiently small and $\Lambda \nearrow \frac{1}{4r_+^2} - c_0$ as ${\epsilon^*, \delta^* \searrow 0}$.  Then since $\epsilon < \frac{1}{4r_+^2} - c_0$, there exist $\epsilon_1$, $\delta_1 > 0$ such that $\Lambda(\epsilon_1, \delta_1) > \epsilon$. Without loss of generality, we may assume that $\delta_1 \leq \delta$, and henceforth we restrict our attention to the (possibly) smaller region $\mathcal{W}(\delta_1)$, using $\mathcal{W}$ to denote it rather than $\mathcal{W}(\delta)$.  We make use of $\epsilon_1$ in what follows.

Now, from condition \Aprimecond$^\prime$ it follows immediately that condition \Acond\, holds in $\mathcal{W} \cap \overline{\mathcal{R}}$.  This in turn implies that condition \Acond\, holds in $\mathcal{A} \cap \mathcal{W}$. (To see that $\mathcal{A}\setminus\del\mathcal{R} = \emptyset$, observe that a point $p \in \mathcal{A}\setminus\del\mathcal{R}$ would have to have a neighborhood lying entirely in $\mathcal{T} \cup \mathcal{A}$; hence by Proposition \ref{handy}, any connected component of $\mathcal{A}\setminus\del\mathcal{R}$ must be an outgoing null ray to the past of $p$ and must intersect either $\del \mathcal{R}$ or $\mathcal{C}_{in}$.  The former case contradicts the achronality of $\mathcal{A}$ at points where condition \Acond\, holds (Proposition \ref{achronal}), and the latter
case we can exclude, without any loss of generality, by suitably shrinking $u_0$.)
Thus by Proposition \ref{connected},
either $\mathcal{W} \cap \mathcal{R}$ contains a rectangle $K(u_1,v_1)$ for some $u_1 \in (0,u_0]$, $v_1 \in [v_0, \infty)$, or the spacetime contains a marginally trapped tube $\mathcal{A}$ which is asymptotic to the event horizon.  We will show that the existence of the rectangle in the former case leads to a contradiction and thus conclude that the latter statement is true.  Furthermore, given how Proposition \ref{connected} was proved, i.e. via Lemma \ref{lem1},  we will then know that in fact it is $\mathcal{A} \cap \mathcal{W}$ which is asymptotic to the event horizon, so in particular $\mathcal{A}$ must be achronal with no ingoing null segments (by Proposition \ref{achronal}) and connected (by Lemma \ref{lem1}), proving the theorem.

For the remainder of the proof, we restrict our attention to the region $K(u_1, v_1)$.  (We may assume without loss of generality that $v_1 \geq v^\prime$.)  Define $\kappa = -\frac{1}{4}\Omega^2 (\del_u r)^{-1}$.  Then $- \Omega^{-2} \del_u r = \frac{1}{4\kappa}$.  Now, using equation (\ref{eq1}) and condition \Bcondone, we have
\begin{eqnarray*} 
\del_u \log (- \Omega^{-2} \del_u r) & = & - r (\del_u r)^{-1} T_{uu} \\
& \leq & - c_1 r (\del_u r),
\end{eqnarray*}
so integrating along an ingoing null ray, we have
\begin{eqnarray*}
\log \left( \frac{\kappa(0,v)}{\kappa(u,v)} \right)  & = & \int_0^u \del_u \log (-\Omega^{-2} \del_u r)(\tilde{u},v) d\tilde{u}\\
& \leq & - \int_0^u c_1 r (\del_u r)(\tilde{u},v) d\tilde{u} \\
& = & -\textstyle\frac{c_1}{2} (r^2(u,v) - r^2(0,v)) \\
& \leq & -\textstyle\frac{c_1}{2} ((r_+- \delta_1)^2 - r_+^2) \\
& \leq & c_1 r_+ \delta_1,
\end{eqnarray*}
which yields
\[ \kappa(u,v) \geq \kappa(0,v) e^{-c_1 r_+ \delta_1}. \]
Then since condition \Ccond\, implies that $\kappa(0,v) \geq \frac{1}{4c_2}$ for all $v \geq v_1$, we have a lower bound 
\[ \kappa \geq \kappa_0 := \textstyle\frac{1}{4c_2} e^{-c_1 r_+ \delta_1} > 0 \] in all of $K(u_1, v_1)$.

\medskip

Next, since $r(0,v)\rightarrow r_+$ as $v \rightarrow \infty$, $\del_v r$ cannot have a positive lower bound along $\mathcal{C}_{out}$; thus there must exist some $V \geq v_1$ such that $\del_v r(0,V) < \epsilon_1$.  By continuity, there is a neighborhood of $(0,V)$ in $\mathcal{G}(u_0, v_0)$ in which this inequality holds, and in particular, there exists some $0< U \leq u_1$ such that 
\begin{equation} \del_v r(u,V) < \epsilon_1 \text{  for all  } 0 \leq u \leq U. \label{rveps}\end{equation}
Now, since $\Lambda(\epsilon_1, \delta_1) > \epsilon$, we  have
\[  
\frac{\delta_1}{r_+}(2c_0 + 3M) + \epsilon < \left( \frac{r_+ - \delta_1}{r_+}   \right)^3 \left(\frac{1}{4r_+^2} \left(1- 2\epsilon_1 c_2 e^{c_1 r_+ \delta_1}\right) - c_0 \right) , \]
or equivalently, setting $r_0 = r_+ - \delta_1$,
\[ 2\delta_1 r_+^2(2c_0 + 3M) + 2r_+^3\epsilon  < \frac{r_0^3}{2}  \left(\frac{1}{r_+^2} \left(1- \frac{\epsilon_1}{2\kappa_0} \right) - 4c_0 \right).  
\] We can thus fix constants $\alpha_0$ and $\alpha_1$ such that
\begin{equation} 2\delta_1 r_+^2(2c_0 + 3M) + 2r_+^3\epsilon  < \alpha_0 < \frac{r_0^3}{2}  \left(\frac{1}{r_+^2} \left(1- \frac{\epsilon_1}{2\kappa_0} \right) - 4c_0  \right)
\label{alpha0}\end{equation}
and 
\begin{equation} \alpha_1 = \alpha_0 - 2\delta_1 r_+^2(2c_0 + 3M) - 2r_+^3\epsilon > 0. \label{alpha1}\end{equation}

As in the proof of Proposition \ref{achronal}, define a function $\alpha$ on $\mathcal{G}(u_0, v_0)$:
\[ \alpha(u,v) = m - 2r^3\,\Omega^{-2}T_{uv}\,(u,v). \]
Using our lower bound for $\kappa$, (\ref{rveps}), (\ref{alpha0}), and condition \Aprimecond$^\prime$, we see that $\alpha > \alpha_0$ on $[0,U] \times \{ V \}$:
\begin{eqnarray*}\alpha & = & m - 2r^3\,\Omega^{-2}T_{uv} \\
& = & \frac{r}{2}\left(1+4\,\Omega^{-2}\, \del_u r \, \del_v r\right) - 2r^3\,\Omega^{-2}T_{uv} \\
& = & \frac{r^3}{2}\left(\frac{1}{r^2}\left(1-\frac{1}{2\kappa} \del_v r\right) - 4\,\Omega^{-2}T_{uv}\right) \\
& \geq & \frac{r_0^3}{2}\left(\frac{1}{r_+^2}\left(1-\frac{\epsilon_1}{2\kappa_0}\right) - 4c_0\right) \\
& > & \alpha_0.
\end{eqnarray*}
Our goal now is to deduce that $\alpha > \alpha_1$ in $K(U,V)$, using \Bcondtwo.
First we compute:
\begin{eqnarray} \del_v\alpha & = & \del_v(m - 2r^3\,\Omega^{-2}T_{uv}) \nonumber\\
& = & \del_v m -\del_v(2r^3\,\Omega^{-2}T_{uv}) \nonumber\\
& = & 2 r^2 \Omega^{-2} (T_{uv} \del_v r - T_{vv} \del_u r)  - 2r^3\del_v(\Omega^{-2}T_{uv}) -6r^2 (\del_v r )\Omega^{-2}T_{uv}\nonumber \\
& = & - 4 r^2 \Omega^{-2} T_{uv} \del_v r - 2 r^2 \Omega^{-2}  T_{vv} \del_u r  - 2r^3\del_v(\Omega^{-2}T_{uv}). \nonumber \\
& \geq & - 4 r^2 \Omega^{-2} T_{uv} \del_v r - 2r^3\del_v(\Omega^{-2}T_{uv}), \label{alphav}
\end{eqnarray}
where the last inequality follows from assumptions \DEC\, and \drduneg.
The next step is to integrate (\ref{alphav}) along an outgoing null ray $\{ u \} \times [V, v)$, but first let us consider the two summands on the right hand side
separately.  First,
\begin{eqnarray*} 
\int_V^v - 4 r^2 \Omega^{-2} T_{uv} \del_v r & = & \int_V^v - \textstyle\frac{4}{3}(\del_v  r^3) (\Omega^{-2} T_{uv}) \\
& \geq & \int_V^v - \textstyle\frac{4}{3} c_0  (\del_v r^3)  \\
& > &  - \textstyle\frac{4}{3} c_0  (r_+^3 - (r_+-\delta_1)^3)  \\
& > &  - 4c_0  r_+^2\delta_1. 
\end{eqnarray*}
For the second summand of (\ref{alphav}), we use the following notation:
given a function $f$,  $f^+ = \max\{ f, 0 \}$ and $f^- = \max\{ -f, 0 \}$, so that $f = f^+ - f^-$.
Then:
\begin{eqnarray*} \displaystyle\int_V^v 2r^3\del_v(\Omega^{-2}T_{uv}) & = & \displaystyle\int_V^v 2r^3[ \del_v(\Omega^{-2}T_{uv}) ]^+ - 2r^3[ \del_v(\Omega^{-2}T_{uv}) ]^- \\
& \leq & \displaystyle\int_V^v 2r_+^3[ \del_v(\Omega^{-2}T_{uv}) ]^+ - 2(r_+-\delta_1)^3[ \del_v(\Omega^{-2}T_{uv}) ]^- \\
& = & \displaystyle\int_V^v 2r_+^3 (\del_v(\Omega^{-2}T_{uv}) )  \\
&& \hspace*{.2in} + \displaystyle\int_V^v  2(3r_+^2\delta_1 - 3r_+ \delta_1^2 + \delta_1^3)[ \del_v(\Omega^{-2}T_{uv}) ]^- \\
& \leq & 2r_+^3 \epsilon +  6 r_+^2\delta_1 \displaystyle\int_V^v [ \del_v(\Omega^{-2}T_{uv}) ]^- \\
& \leq & 2r_+^3 \epsilon +  6 r_+^2\delta_1 \displaystyle\int_V^v | \del_v(\Omega^{-2}T_{uv}) | \\
& \leq & 2r_+^3 \epsilon +  6 r_+^2\delta_1 M.
\end{eqnarray*}
Integrating (\ref{alphav}) now yields
\begin{eqnarray*} \alpha(u,v) & > & \alpha(u,V) - 4c_0  r_+^2\delta_1 - (2r_+^3 \epsilon +  6 r_+^2\delta_1 M) \\
&  > &  \alpha_0 - 2r_+^3 \epsilon - 2r_+^2\delta_1 ( 2c_0   +  3  M ) \\
& = & \alpha_1.
\end{eqnarray*}
Thus we conclude that $\alpha > \alpha_1$ in all of $K(U,V)$.

\medskip  
Finally, recall from equation (\ref{druv}) that
\[ \del^2_{uv} r =  -\textstyle\frac{1}{2}\Omega^2 r^{-2} \alpha, \]
or, using the definition of $\kappa$, 
\begin{equation*}
\del^2_{uv} r =  2  \kappa r^{-2} (\del_u r)\alpha.
\end{equation*}
Rearranging and applying our bounds in our region of interest, we have
\begin{eqnarray*} \del_v \log (-\del_u r) & = & 2 \kappa r^{-2} \alpha \\
& > & 2 \kappa_0 r_+^{-2} \alpha_1 ,
\end{eqnarray*}
and so integrating along an outgoing ray yields
\[  \frac{\del_u r(u,v)}{\del_u r(u,V)} > e^{ 2 \kappa_0 r_+^{-2} \alpha_1 (v-V)}, \]
and hence
\begin{equation}\label{forcor}  - \del_u r(u,v) > - \del_u r(u,V) \,e^{2 \kappa_0 r_+^{-2} \alpha_1 (v-V)}. \end{equation}
Assume $\del_u r(u,V) \leq -b_0 < 0$ for all $0 \leq u \leq U$, let
\[ b_1 = 2 \kappa_0 r_+^{-2} \alpha_1 , \]
and set
\[ b_2 = b_0 e^{-b_1 V}; \]
then
\[ - \del_u r(u,v) > b_2 e^{b_1 v} \]
and so integrating along an ingoing null ray, we get
\[ r(0,v) - r(u,v) > b_2 e^{b_1 v} u, \]
i.e.
\[ r(u,v) < r(0,v) - b_2 e^{b_1 v} u. \]
But for any $u > 0$, the right-hand side tends to $-\infty$ as $v \rightarrow \infty$, while the left-hand side is positive.
Thus we have arrived at a contradiction, so no such rectangle $K(U,V)$ can be contained in $\mathcal{R}$, and the statement of the theorem follows. 
\end{proof}

\noindent \it Remark: \rm  
Given the last estimate obtained in the proof of Theorem \ref{mainthm}, it is essentially immediate that the event horizon of the black hole is future geodesically complete, i.e.\ $\mathcal{C}_{out}$ has infinite affine length, whenever the hypotheses of Theorem \ref{mainthm} are satisfied. 
(Note, however, that this result does not \it require \rm all of the hypotheses of Theorem \ref{mainthm}; the argument below could be refined such that  
assumptions  \Ccond\, and \Aprimecond$^\prime$ (restricted to $\mathcal{W} \cap \mathcal{C}_{out}$) are sufficient.)

Suppose $s$ is an affine parameter for $\mathcal{C}_{out} = \{ 0 \} \times [ v_0, \infty) = \{ (0, v(s)) \}$ which increases to the future. Then the vector field
$X = \frac{dv}{ds} \frac{\del}{\del v}$ satisfies
$\nabla_{X} X = 0$, which in this setting becomes 
\[ \textstyle \frac{dv}{ds} \left[\del_v\left(\frac{dv}{ds}\right) + \left(\frac{dv}{ds}\right)\left(\del_v(\log \Omega^2)\right)\right] = 0, \]
or equivalently, since $\frac{dv}{ds} > 0,$
\[ \del_v\left(\log(\Omega^2 \textstyle\frac{dv}{ds})\right) = 0. \]
Integrating, we have
\[ \textstyle\frac{dv}{ds} = a_0 \Omega^{-2}\]
for some $a_0 > 0$,
and so now taking $s$ as a function of the outgoing null coordinate $v$, we have
\[ s(v) = a_1 + a_2 \int_{v_0}^{v} \Omega^2 (0, \overline{v}) d\overline{v},\]
some $a_1, a_2 \in \mathbb{R}$, $a_2 > 0$.
Thus $s$ has infinite range if and only if $\Omega \notin L^2([v_0, \infty))$.

Now, one of the hypotheses of Theorem \ref{mainthm} was that $(-\del_u r)\Omega^{-2} \leq c_2$ along $\mathcal{C}_{out} \cap \mathcal{W}$ for some constant $c_2  > 0$ (condition \Ccond), and in the proof of the theorem, we found that for some $U > 0$, $V \geq v_0$ and some $b_1, b_2 > 0$,  
\[- \del_u r(u,v) > b_2 e^{b_1 v} \] 
for all $0 \leq u \leq U$, $V \leq v$ (equation (\ref{forcor})).  Putting these inequalities together and evaluating along $\mathcal{C}_{out}$, we have
\[ b_2 e^{b_1 v} < - \del_u r(0,v) \,\leq \, c_2 \Omega^2(0,v)\]
for all $V \leq v$, so in fact $\Omega \notin L^2([v_0, \infty))$ and $\mathcal{C}_{out}$ is future complete.

%%%%%%%%%%%%%%%%%%%%%%%%%%%%%%%%%%%%%%%%%%%%%%%%%%%%%%%%%%%%%%%%%%%%%%%%%%%%%%%%%%%%%%%%%%%%%%%%%%%%%%%%%%%%%%%%%%%%%%%%%%%%%%%
%%%%%%%%%%%%%%%%%%%%%%%%%%%%%%%%%%%%%%%%%%%%%%%%%%%%%%%%%%%%%%%%%%%%%%%%%%%%%%%%%%%%%%%%%%%%%%%%%%%%%%%%%%%%%%%%%%%%%%%%%%%%%%%

\section{Immediate applications}\label{aps}
\subsection{Price law decay \& length}

\begin{thm} \label{pricelaw} \,
Suppose $(\mathcal{G}(u_0, v_0), -\Omega^2\,du\,dv)$ is a spacetime obtained as in Section \ref{ssivp} with radial function $r$, and suppose it satisfies assumptions \rm \DEC-\extprinc\, \it of Section \ref{bhea}.  Suppose $\mathcal{A}_0$ is a connected component of $\mathcal{A}$ along which condition \rm \Acond\, \it is satisfied, and
suppose that in addition,
\[ T_{vv} \leq c_3 v^{-2 - \epsilon} \text{ along }\mathcal{A}_0 \]
for some $c_3 \geq 0$,  $\epsilon > 0$. Then $\mathcal{A}_0$ has finite length with respect to the induced metric.
\end{thm}
\smallskip

\noindent \it Remark: \rm The rate of decay of $T_{vv}$ given corresponds to that of Price's law; cf. \cite{P,D1}. 
Theorem \ref{pricelaw} applies in particular to the case of the marginally trapped tube $\mathcal{A}\cap \mathcal{W}$ obtained in Theorem \ref{mainthm},  but it does not require that the tube terminate at $i^+$ in order to be valid.  In the context of Theorem \ref{mainthm}, this decay rate gives a direct measure of how quickly the tube approaches the event horizon.  

\medskip

\begin{proof}  Since $\mathcal{A}_0$ is connected and achronal with no ingoing null segments, we may parameterize it by its $v$ coordinate,  i.e. $\gamma(v) = (u(v), v) \in \mathcal{A}_0$.  If the domain of the $v$ coordinate is bounded, then the result is trivially true, so suppose $v$ has domain $[V, \infty)$ for some $V \geq v_0$.  Then we have
\[ |\dot{\gamma}(v)|^2 = \langle \dot{\gamma}(v), \dot{\gamma}(v) \rangle = -\Omega^2 (\textstyle\frac{du}{dv}). \]
Using the relation $\dot{\gamma}(\del_v r) = 0$, we readily compute that
\[ \frac{du}{dv} = - \displaystyle\frac{\del_{vv}^2 r}{\del_{uv}^2 r} = -\frac{2r^3 T_{vv}}{\Omega^2 \alpha}, \]
so
\[ |\dot{\gamma}(v)|^2 = \frac{2r^3 T_{vv}}{\alpha}. \]
As in the proof of Proposition \ref{achronal}, we compute that $\alpha > \alpha_0$ along $\mathcal{A}$, so setting $b_3 = \sqrt{2r_+^3\alpha_0^{-1}}$, we have
\[ \int_V^\infty |\dot{\gamma}(v)| dv \leq \int_V^\infty b_3 ({T_{vv}(\gamma(v))})^{\frac{1}{2}} dv \leq \int_V^\infty b_3 c_3 v^{-1 - \epsilon/2} dv < \infty, \] 
i.e., the length of $\mathcal{A}_0$ is finite. 
\end{proof}

\subsection{Vaidya spacetimes} \label{Vaidyapp}

\medskip
Perhaps the simplest example at hand when one is working with dynamical horizons is that of the (ingoing) Vaidya spacetime, the spherically symmetric solution to Einstein's equations with an ingoing null fluid as source.  It is widely accepted that the Vaidya marginally trapped tube is asymptotic to the event horizon, but the literature seems to be lacking an analytical proof of this behavior for an arbitrary mass function, so it is worth seeing how our results apply to this case.

Recall that the ingoing Vaidya metric is given in terms of the ingoing Eddington-Finkelstein coordinates  $(v, r, \theta, \varphi)$ by
\[ g = -\left(1 - \frac{2M(v)}{r}\right)dv^2 + 2dvdr + r^2 (d\theta^2 + \sin^2 \theta d\varphi^2), \]
with stress-energy tensor
\[ T = \frac{\dot{M}(v)}{r^2} dv^2, \]
where and $M(v)$ is any smooth function of $v$.  One can show directly that the marginally trapped tube is the hypersurface at which $r = 2M(v)$; it is spacelike where $\dot{M}(v) > 0$, null where $\dot{M}(v) = 0$, and timelike where $\dot{M}(v) < 0$.

By inspection $T$ satisfies the dominant energy condition, assumption \DEC, if and only if $M(v)$ is nondecreasing.  
We restrict our attention to a characteristic rectangle in which $M$ is strictly positive and indicate how the remaining assumptions \past-\extprinc\, are satisfied:
the metric is regular everywhere in the rectangle except at the singularity at $r = 0$ (albeit not in the coordinates given above), so it follows that assumption \extprinc\, is satisfied, and furthermore, that singularity is evidently spacelike, so \past\, is satisfied as well.
The inner expansion of each round 2-sphere is (some positive multiple of) $-\frac{2}{r}$, i.e. it is  strictly negative, so assumption \drduneg\, holds.  Finally, assuming $M(v) < M_0$ for some limiting value $M_0 < \infty$, assumptions \rbounds, \mpos, and \drdvpos\, are all satisfied as well (the strict inequality is what yields \drdvpos).

Now, in order to check the hypotheses of Theorem \ref{mainthm}, it seems we ought to convert from Eddington-Finkelstein to double-null coordinates.  Unfortunately, to make such a conversion explicitly is impossible in general; see \cite{WL}.  However, we can still compute the relevant quantities from first principles.   Following the treatment given in \cite{WL}, we find
that in double-null coordinates $(u, v, \theta, \varphi)$  with $v$ scaled such that its domain is $[v_0, \infty)$, the only nonzero component of $T_{\alpha \beta}$ is $T_{vv}$.  Thus conditions \Aprimecond$^\prime$, \Bcondone, and \Bcondtwo\, are all trivially satisfied.  We can also deduce that the term $-(\del_u r)\Omega^{-2} = \frac{1}{2}$ everywhere, so \Ccond\, is satisfied as well.  Thus we conclude from Theorem \ref{mainthm} that the marginally trapped tube is asymptotic to the event horizon.  Furthermore, as one might expect, Theorem \ref{pricelaw} tells us that the length of the tube depends on the rate of decay of $\dot{M}(v)$ as $v \rightarrow \infty$.  In particular,  if $\dot{M}(v) = O(v^{-2-\epsilon})$ for some $\epsilon > 0$, or more generally, if $(\dot{M}(v))^{1/2} \in L^1([V, \infty))$ for some $V$, then the tube has finite length. 

%%%%%%%%%%%%%%%%%%%%%%%%%%%%%%%%%%%%%%%%%%%%%%%%%%%%%%%%% HIGGS FIELD %%%%%%%%%%%%%%%%%%%%%%%%%%%%%%%%%%%%%%%%%%%%%%%%%%%%%%%%%%%%%%%%%%%%%

\subsection{Self-gravitating Higgs fields}
\medskip
\noindent

\medskip
We next consider a self-gravitating Higgs field with non-zero potential.  This matter model consists of a scalar function $\phi$ on the spacetime and a potential function $V(\phi)$ such that
\begin{equation}\label{sfeq} \Box\, \phi = g^{\alpha \beta}\phi_{;\alpha \beta}  = V^\prime(\phi). \end{equation}
The stress-energy tensor then takes the form
\begin{equation}\label{sfT} T_{\alpha \beta} = \phi_{; \alpha} \phi_{; \beta} - \left( \textstyle\frac{1}{2} \phi_{;\gamma} \phi^{;\gamma} + V(\phi) \right) g_{\alpha \beta}. \end{equation} 
In our  spherically symmetric setting, $\phi = \phi(u, v)$, so the evolution equation (\ref{sfeq}) becomes
\begin{equation}\label{sfequv} V^\prime(\phi) = -4\Omega^{-2} ( \del^2_{uv}\phi + \del_u\phi\,(\del_v \log r) + \del_v\phi\,(\del_u \log r)) \end{equation}
and in double-null coordinates, (\ref{sfT}) yields
\begin{equation*} T_{uu} = (\del_u\phi)^2 \end{equation*}
\begin{equation*} T_{vv} = (\del_v\phi)^2 \end{equation*}
and
\begin{equation*} T_{uv} = \textstyle\frac{1}{2}\Omega^2 V(\phi). \end{equation*}
Note that the dominant energy condition (\DEC) is satisfied if and only if $V(\phi)$ is nonnegative.  The extension principle (\extprinc) is known to hold for self-gravitating Higgs fields if $V \geq -C$ for any finite $C$ \cite{D3}.

We give two applications of Theorem \ref{mainthm} to self-gravitating Higgs field black hole spacetimes.  
In Theorem \ref{HiggsR}, we assume that the scalar field and the potential both satisfy weak Price-law-like decay conditions on the event horizon, namely that $|\del_v \phi|$ and $|V^\prime(\phi)| \in O(v^{-p})$ for some constant $p >\frac 12$.
For Theorem \ref{HiggsM}, we make only certain monotonicity and smallness assumptions, including that $V$ is convex.  
In both cases, we extract the hypotheses of Theorem \ref{mainthm} and conclude that the respective black holes each contain a marginally trapped tube which is asymptotic to the event horizon, achronal with no ingoing null segments, and connected for large $v$.  

The advantage of Theorem \ref{HiggsM} over Theorem \ref{HiggsR} is that it does not require an explicit rate of decay.  
However, the assumptions which we must make in the absence of such a decay rate are highly nontrivial.  
In particular, in the case of a Klein-Gordon potential of mass $\overline{m}$, i.e. $V(\phi) = \frac{1}{2}\overline{m}^2 \phi^2$, the hypotheses of Theorem \ref{HiggsM} are satisfied only if $\phi$ decays exponentially along the event horizon.  
In the case of an exponential potential $V(\phi) = ce^{k\phi}$, the hypotheses cannot even be satisfied simultaneously.  
By contrast, Theorem \ref{HiggsR} can be applied in both such settings.  
On the other hand, for any potential of the form $V(\phi) = c \phi^{k+2}$, $k > 0$, the hypotheses of Theorem \ref{HiggsM} may be readily satisfied; in this case they imply  that $\phi \in O(v^{-\frac{1}{k}})$ and $V^\prime(\phi) \in O(v^{-1-\frac{1}{k}})$ along the event horizon but make no \it a priori \rm restriction on $|\del_v \phi|$.  
Indeed, it is possible to construct an admissible $\phi$ along the event horizon in this setting such that $\limsup_{v\rightarrow\infty}|\del_v \phi| v^p = \infty$ for any $p > 0$, which then implies that Price law decay per se does not hold.   

\medskip

For both of the following theorems, we assume we have initial data $r$, $\Omega$, $\phi$ for a self-gravitating Higgs field along the null hypersurfaces $\mathcal{C}_{in} \cup \mathcal{C}_{out} = [0,u_0] \times \{ v_0 \} \cup \{ 0 \} \times [v_0, \infty)$ with nonnegative potential function $V \in C^2(\mathbb{R})$,  and we suppose the data satisfy assumptions \rbounds-\drdvpos,
namely: $r\leq r_+$ and $0 \leq m \leq m_+$ along $\mathcal{C}_{in} \cup \mathcal{C}_{out}$, and $\del_ur < 0$, $\del_v r > 0$ along $\mathcal{C}_{out}$.

%%%%%%%%%%%%%%%%%%%%%%%%%%%%%%%%%%%%%%%%%%%%%%%%%%%%%%%%%%%%%%%%%%%%%%%%%%%%%%%%%
%%%%%%%%%%%%%%%%%%%%%%%%%%%%%%%%%%%%%%%%%%%%%%%%%%%%%%%%%%%%%%%%%%%%%%%%%%%%%%%%%
%										%
%										%
%				Price law version				%
%										%
%										%
%%%%%%%%%%%%%%%%%%%%%%%%%%%%%%%%%%%%%%%%%%%%%%%%%%%%%%%%%%%%%%%%%%%%%%%%%%%%%%%%%
%%%%%%%%%%%%%%%%%%%%%%%%%%%%%%%%%%%%%%%%%%%%%%%%%%%%%%%%%%%%%%%%%%%%%%%%%%%%%%%%%

\begin{thm}\label{HiggsR}
Fix a constant $p > \frac 12$ and a function $\eta(v) > 0$ such that $\eta(v)$ decreases monotonically to $0$ as $v$ tends to infinity.
Suppose the second derivative of the potential $V$ is bounded, i.e. there exists a constant $B$ such that
\[ |V^{\prime\prime}(x)| \leq B \]
on the interval $(\phi_0 -  \delta_0, \,\phi_1 + \delta_0)$ for some $\delta_0 > 0$, where $\phi_0$ and $\phi_1$ are the (possibly infinite) $\inf$ and $\sup$ of $\phi$ along $\mathcal{C}_{out}$, respectively.
If along $\mathcal{C}_{out}$ the initial data satisfy
\[ \del_v r < \eta(v), \]
\[ |\del_v \phi | < \textstyle \frac 12 b_1 v^{-p}, \]
\[ | V^\prime(\phi) | < \textstyle \frac 12  b_2 v^{-p}, \]
\begin{equation}\label{c2c3choice} c_3 \leq -(\del_u r)\Omega^{-2}\leq \textstyle\frac 12 c_2, \end{equation}
and
\begin{equation}\liminf_{v\rightarrow\infty} V(\phi) < \frac{1}{4r_+^2}, \label{liminfV} \end{equation}
for some positive constants $b_1$, $b_2$, $c_2$, and $c_3$, 
then the result of Theorem \ref{mainthm} holds for maximal development of these initial data.
\end{thm}

%%%%%%%%%%%%%%%%%%%%%%%%%%%%%% REMARKS %%%%%%%%%%%%%%%%%%%%%%%%%%%%%%%%%%%%%%%%%%%
\smallskip

\noindent \it Remark: \rm  Note that if the constant $p > 1$, then $\del_v\phi$ is integrable along the event horizon, which in turn implies that the domain of $\phi$ is compact and hence that $V^{\prime\prime}$ is \it a priori \rm bounded on the relevant domain.  Thus the hypothesis that $V^{\prime\prime}$ be bounded is only necessary for $\frac{1}{2} < p \leq 1$.

\medskip

%%%%%%%%%%%%%%%%%%%%%%%%%%%%%%%%%%%%%%%%%%%%%%%%%%%%%%%%%%%%%%%%%%%%%%%%%%%%%%%%%
%%%%%%%%%%%%%%%%%%%%%%%%%%%%%%%%%%%%%%%%%%%%%%%%%%%%%%%%%%%%%%%%%%%%%%%%%%%%%%%%%
%										%
%										%
%			Monotonicity version					%
%										%
%										%
%%%%%%%%%%%%%%%%%%%%%%%%%%%%%%%%%%%%%%%%%%%%%%%%%%%%%%%%%%%%%%%%%%%%%%%%%%%%%%%%%
%%%%%%%%%%%%%%%%%%%%%%%%%%%%%%%%%%%%%%%%%%%%%%%%%%%%%%%%%%%%%%%%%%%%%%%%%%%%%%%%%

\begin{thm}\label{HiggsM}
Suppose there exist positive constants $c_0$, $c_1$, $c_2$, $c_3$, and $c_4$ such that
\begin{equation}\label{c0choice} c_0 < \frac{1}{4r_+^2}; \end{equation}
along $\mathcal{C}_{in}$
\begin{equation}\label{c1choice} \displaystyle \left( \frac{\del_u \phi}{\del_u r} \right)^{\!2}  <   c_1;\end{equation}
and along $\mathcal{C}_{out}$
\begin{equation}\label{c2c3choiceM} c_3 \leq -(\del_u r)\Omega^{-2}\leq \textstyle\frac 12 c_2 \end{equation}
and
\begin{equation}\label{c4choice} V^\prime(\phi) < c_4 |\del_v \phi |. \end{equation} 
Suppose also that the potential $V$ satisfies
\begin{equation}\label{Bchoice} 0 \leq \,V^{\prime\prime}(x) \leq B
\end{equation}
on the interval $(\phi_0 -  \delta_0, \,\phi_1)$ for some $\delta_0 > 0$, where $\phi_0$ and $\phi_1$ are the (possibly infinite) $\inf$ and $\sup$ of $\phi$ along $\mathcal{C}_{out}$, respectively, and $B$ is a constant satisfying $B <r_+^{-2}$. 

\smallskip 
If along $\mathcal{C}_{out}$ the initial data satisfy
\[ \del_v r < \epsilon \]
\[ V(\phi) < \textstyle \frac 12 \epsilon^{\prime} \]
\[ |\del_v \phi | < \textstyle \frac 12 \epsilon^{\prime\prime} \]
for sufficiently small $\epsilon$, $\epsilon^\prime$, and $\epsilon^{\prime\prime} > 0$, as well as
\begin{eqnarray*} 
\del_v\phi & < & 0 
\\ \del_u\phi & < & 0 
\\ V^\prime (\phi) - 4\sqrt{c_1}c_2 (\del_v \log r) - 4c_3r_+^{-1}(\del_v \phi) & > & 0, 
\end{eqnarray*}
and either
\[  V^\prime(\phi) \leq 0 \]
or
\[ |\phi_0| < \infty, \]
then the result of Theorem \ref{mainthm} holds for maximal development of these initial data. 
\end{thm}

%%%%%%%%%%%%%%%%%%%%%%%%%%%%%% REMARKS %%%%%%%%%%%%%%%%%%%%%%%%%%%%%%%%%%%%%%%%%%%

\noindent \it Remarks: \rm  Stated more precisely, the requirement that the constants $\epsilon$, $\epsilon^\prime$, and $\epsilon^{\prime\prime}$ be sufficiently small  is the following:
\[ 2\epsilon^\prime + 8c_2 r_+^{-2}\epsilon < r_+^{-2} - B,\] 
\[ \epsilon^\prime < \min \left\{ \frac{1}{2r_+^2} - 2c_0 , 2c_0 \right\}, \]
and
\[ \epsilon^{\prime\prime} < \frac{\sqrt{c_1}c_3(1-2r_+^2\epsilon^\prime - 4c_2 \epsilon)}{c_2(c_4r_+ + 4c_3)}. \]

The existence of constants $c_0$ and $c_1$ satisfying (\ref{c0choice}) and (\ref{c1choice}) is not restrictive, but that of constants $c_2$ and $c_3$ in (\ref{c2c3choiceM}) is.   Inequalities (\ref{dvphineg}) and (\ref{duphineg}) together imply that the scalar field $\phi$ has a timelike gradient.
Note that if $V^\prime(\phi) \leq 0$ along $\mathcal{C}_{out}$, then (\ref{c4choice}) is trivially true for any $c_4$. 

\bigskip

%%%%%%%%%%%%%%%%%%%%%%%%%%%%% PROOF OF HIGGS_R %%%%%%%%%%%%%%%%%%%%%%%%%%%%%%%%%%%%%%%%%%%%%%%%
%  		(the R stands for "Rate", as in prescribed decay rate)
%
%
\begin{proof}[Proof of Theorem \ref{HiggsR}]
Let $\mathcal{G}(u_0, v_0)$ denote the maximal development of the given initial data.
By (\ref{liminfV}), we may choose a positive constant $c_0$ such that
\[  \liminf_{v\rightarrow\infty} V(\phi(0,v)) < c_0 < \textstyle\frac{1}{4r_+^2}.\]
Thus there exist some small $0 < \epsilon < 2c_0$ and a sequence $\{ \overline{v}_k \} \rightarrow \infty$ such that $V(\phi(0,\overline{v}_k)) < c_0 - \frac 12 \epsilon$ for all $k$.
Choose $\epsilon^\prime > 0$ such that
\[ \epsilon^\prime < \min\{ c_0 - \textstyle\frac 12 \epsilon , \frac{1}{4r_+^2} - c_0 \}. \]
Since $\overline{v}_k \rightarrow \infty$, we can find $K$ sufficiently large that for $v \geq \overline{v}_K$,
\[ \frac{b_1b_2}{2p-1} v^{1-2p} < \epsilon^\prime,\]
\[ 2c_2\eta(v) < r_+^2 \epsilon, \]
and
\begin{equation}\label{v13} 2c_2r_+p\left(\frac{\log v}{v}\right) < r_+^2\epsilon - 2c_2\eta(v) .\end{equation}
Set $v_1 = \overline{v}_K$ and note that by construction $V(\phi(0,v_1)) < c_0 - \frac 12 \epsilon$.

\medskip

Next, set $b_0 = \eta(v_1)$,  let
\begin{eqnarray*} \overline{b_3} & = & 2\left(\frac{b_2}{4c_3}+\frac{2b_1}{r_+}\right) \left(\frac{2c_2r_+}{r_+^2\epsilon - 2c_2b_0}\right)
\\
&& \hspace*{.3in}\cdot 
\left(v_1^{-p} + \left(1- \left(\frac{2c_2r_+p}{r_+^2\epsilon - 2c_2b_0}\right)\frac{\log v_1}{v_1}\right)^{\!\!-p\,}\right),
\end{eqnarray*}
and fix $b_3$ such that
\[ b_3 > \max\left\{ 2 \left| \frac{\del_u \phi}{\del_u r}(0,v_1) \right| v_1^p, \,\,\overline{b_3}  \right\}. \]
Now, continuity at the point $(0,v_1)$ and our initial conditions along $\mathcal{C}_{out}$ imply that there exists $u_1 > 0$ sufficiently small that 
\[ \del_v r(u,v_1) < b_0, \]
\[ |\del_v \phi(u,v_1) | < \textstyle \frac 12 b_1 v_1^{-p}, \]
\[ | V^\prime(\phi(u,v_1)) | < \textstyle \frac 12  b_2 v_1^{-p}, \]
\[ \left| \frac{\del_u \phi}{\del_u r}(u,v_1) \right| < \textstyle \frac 12 b_3 v_1^{-p}, \]
and
\[ V(\phi(u,v_1)) < c_0 - \textstyle\frac 12 \epsilon \]
for all $u \in [0, u_1]$.
Set $\mathcal{C}_{in}^\prime = [0, u_1] \times \{ v_1 \}$ and $\mathcal{C}_{out}^\prime = \{0 \} \times [v_1, \infty)$.
Henceforth we will consider the subregion of $\mathcal{G}(u_0,v_0)$ given by
\[ \mathcal{G}(u_1,v_1) := K(u_1, v_1) \cap \mathcal{G}(u_0,v_0), \]
i.e. the maximal development of the induced initial data on 
$\mathcal{C}^\prime_{in} \cup \mathcal{C}^\prime_{out}$.  

\medskip
Finally, we choose 
\[ 0 < \delta < \min\left\{ \frac{r_+}{2}, \, \displaystyle\frac{b_1}{2}\left( \displaystyle\frac{b_2}{4c_3} + \frac{2b_0 b_3 + 2b_1}{r_+} \right)^{\!-1},  \,\displaystyle\frac{b_2}{2Bb_3}, \, \frac{\delta_0 v_1^p}{b_3}, \, \frac{v_1^{2p}}{2b_3^2 r_+} \right\},\]
and as usual, define
\[ \mathcal{W} = \{ (u,v) \in \mathcal{G}(u_1, v_1) \,\,|\,\, r(u,v) \geq r_+ - \delta  \}  .\]

Define a region $\mathcal{V}$ as the set of points $(u,v) \in  \mathcal{G}(u_1, v_1)$ such that the following seven inequalities hold for all $(\tilde{u}, \tilde{v}) \in J^- (u,v) \cap \mathcal{G}(u_1, v_1)$: 
\begin{eqnarray}
\del_v r & < & b_0  \label{b0} \\
|\del_v \phi | & < & b_1 v^{-p} \label{b1}\\
| V^\prime(\phi) | & < & b_2 v^{-p} \label{b2}\\
\left|\frac{\del_u \phi}{\del_u r}\right| & < & b_3 v^{-p} \label{b3}\\
-(\del_u r)\Omega^{-2}& < & c_2 \label{c2}\\ 
V(\phi) & < & 2c_0 - \epsilon \label{Vbound}\\
\del_v r & > & 0. \label{reg}
\end{eqnarray}
Note that (\ref{reg}) implies that $\mathcal{V} \subset \mathcal{R}$, so $\mathcal{W} \cap \mathcal{V} \subset \mathcal{W} \cap \mathcal{R}$.  Since we can easily extract the hypotheses of Theorem \ref{mainthm} from these inequalities, our goal is to prove that in fact $\mathcal{W} \cap \mathcal{R} \subset \mathcal{V}$.  We accomplish this by means of a bootstrap argument showing that $\mathcal{W} \cap \mathcal{R} = \mathcal{W} \cap \mathcal{V}$.
Before proceeding, however, let us show that the hypotheses of Theorem \ref{mainthm} are satisfied.  

First we check that conditions \Aprimecond$^{\!\prime}$-\Ccond\, hold in $\mathcal{V}$.  Note that $\mathcal{V}$ is a past set by definition, and our choices of $u_1$ and $v_1$ imply that $\mathcal{C}^\prime_{in} \subset \mathcal{V}$.  Thus for $(u,v) \in \mathcal{V}$, we have
\begin{equation} \int_{v_1}^v |\del_v V(\phi (u,\tilde{v})|\,d\tilde{v} \leq b_1 b_2 \int_{v_1}^v \tilde{v}^{-2p} \,d\tilde{v} < \frac{b_1 b_2}{2p-1} v_1^{1-2p} < \epsilon^\prime
.\label{intV}\end{equation}
Since $\epsilon^\prime < \frac{1}{4r_+^2} - c_0$, (\ref{intV}) implies that \Bcondtwo\, is satisfied, and since $\epsilon^\prime < c_0 - \textstyle\frac 12 \epsilon$, it also implies that
\[ \textstyle\frac{1}{2} V(\phi(u,v)) < \frac{1}{2}V(\phi(u,v_1)) + \frac{1}{2}\epsilon^\prime < c_0 - \textstyle\frac 12 \epsilon < c_0 < \frac{1}{4r_+^2},\]
so \Aprimecond$^\prime$ is satisfied as well.
Condition \Bcondone\, follows immediately from (\ref{b3}), and
\Ccond\, follows from the hypothesis (\ref{c2c3choice}).  

We must also verify that assumptions \DEC-\extprinc\, hold in $\mathcal{G}(u_1, v_1)$.
The requirement that $V$ be nonnegative implies \DEC, while \past\, follows by construction, since $\mathcal{G}(u_1, v_1)$ is the maximal development of initial data on $\mathcal{C}^\prime_{in} \cup \mathcal{C}^\prime_{out}$.  Assumptions \rbounds\, and \mpos\,  hold on $\mathcal{C}^\prime_{in} \cup \mathcal{C}^\prime_{out}$ by monotonicities of $r$ and $m$ in $\mathcal{R}$, respectively, while \drduneg\, and \drdvpos\, were among the hypotheses of the theorem.  Finally, assumption \extprinc\, holds by \cite{D3} since $V$ is nonnegative.  

\medskip

We now turn to the bootstrap argument, which we carry out as follows:  we first retrieve (strict) inequalities (\ref{b3}) and (\ref{Vbound}) in $\overline{\mathcal{V}}$, where the set closure is taken with respect to $\mathcal{G}(u_1, v_1)$, i.e. $\overline{\mathcal{V}} = \overline{\mathcal{V}} \cap \mathcal{G}(u_1, v_1)$.  Since
inequalities (\ref{b0})-(\ref{b2}), (\ref{c2}) and (\ref{reg}) hold along $\mathcal{C}^\prime_{out}$ by hypothesis, a continuity argument then implies that $\mathcal{C}^\prime_{out} \subset \mathcal{V}$, i.e. that both (\ref{b3}) and (\ref{Vbound}) hold along all of $\mathcal{C}^\prime_{out}$.  Thus $\mathcal{W} \cap \mathcal{V} \neq \emptyset$, since $\mathcal{W}$ must contain a neighborhood of $i^+ = (0,\infty)$.  We then retrieve inequalities (\ref{b0})-(\ref{b2}) and (\ref{c2}) in $\overline{\mathcal{W} \cap \mathcal{V}}$ and conclude, again by continuity, that in fact $\mathcal{W} \cap \mathcal{V} = \mathcal{W} \cap \mathcal{R}$.  

\medskip

It will again be convenient to use the quantity $\kappa$ introduced in the proof of Theorem \ref{mainthm},
\begin{equation}\label{kappadef} \kappa = -\textstyle\frac 14 \Omega^2 (\del_u r)^{-1} = \displaystyle\frac{\del_v r}{1-\frac{2m}{r}}.\end{equation}
Equation (\ref{eq1}) implies that $\del_u \kappa \leq 0$, and combining this fact with (\ref{c2c3choice}), we have
$ \kappa \leq \frac{1}{4c_3}$ in all of $\mathcal{G}(u_1, v_1)$.  The bootstrap inequality (\ref{c2}) implies that
$ \kappa \geq \frac{1}{4c_2}$ in all of $\overline{\mathcal{V}}$ as well,
so by (\ref{b0}),
\begin{equation} \textstyle(1-\frac{2m}{r})(u,v) = (\del_v r) \kappa^{-1}(u,v) \leq 4 c_2 b_0 \label{2mr_Rbound}\end{equation}
in $\overline{\mathcal{V}}$.
Also, note that combining equations (\ref{eq1}) and (\ref{eq3}) (or alternately (\ref{eq2}) and (\ref{eq4})) yields
\begin{equation}\label{ruv} \del^2_{uv} r = \textstyle\frac{1}{4} \Omega^2 r^{-1} \left( 2 r^2 V(\phi)  + (1-\frac{2m}{r}) -1 \right).
\end{equation}

\medskip

Let us now retrieve inequality (\ref{b3}) in $\mathcal{V}$. 
First we observe that (\ref{sfequv}) may be rearranged as
\begin{equation}\label{phiuv} \del^2_{uv}\phi = -\textstyle\frac{1}{4}\Omega^{2} V^\prime(\phi) - (\del_u\phi)(\del_v \log r) - (\del_v\phi)(\del_u \log r).   
\end{equation}
Set $r_1 = r(0, v_1)$ and note that $\del_v r > 0$ implies that $r \geq r_1$ on all of $\mathcal{C}_{out}$.  
We compute:
\begin{eqnarray} \del_v \left( \frac{\del_u \phi }{ \del_u r } \right) 
& = & \frac{\del^2_{uv}\phi}{\del_u r} - \left( \frac{ \del_u\phi }{\del_u r}\right) \left( \frac{\del^{2}_{uv}r } {\del_u r } \right) \nonumber\\
& = & \left( -\textstyle\frac{1}{4}\Omega^{2} V^\prime(\phi) - (\del_u\phi)(\del_v \log r) - (\del_v\phi)(\del_u \log r) \right)(\del_u r)^{-1} \nonumber\\
&&
- \left( \frac{ \del_u\phi }{\del_u r}\right) \left( \textstyle\frac{1}{4} \Omega^2 r^{-1} \left[ 2 r^2 V(\phi) - (1-\frac{2m}{r}) -1 \right] \right)( \del_u r)^{-1} \nonumber\\
& = & \kappa V^\prime(\phi) - \left(\frac{\del_u\phi}{\del_u r}\right)(\del_v \log r) - \frac{\del_v\phi}{r} \nonumber\\
&&
- \left( \frac{ \del_u\phi }{\del_u r}\right) \kappa r^{-1} \left[ 1- 2 r^2 V(\phi) - (1-\textstyle\frac{2m}{r}) \right]  \nonumber\\
& = & \kappa V^\prime(\phi)  - \frac{\del_v\phi}{r} \nonumber\\
&&
- \left( \frac{ \del_u\phi }{\del_u r}\right) \left( \kappa r^{-1} \left[ 1- 2 r^2 V(\phi) - (1-\textstyle\frac{2m}{r}) \right] + (\del_v \log r) \right). \nonumber
\end{eqnarray}
Let
\[ 
A := \kappa r^{-1} \left[ 1- 2 r^2 V(\phi) - (1-\textstyle\frac{2m}{r}) \right] + (\del_v \log r), 
\]
so that we may write 
\begin{equation}\del_v \left( \frac{\del_u \phi }{ \del_u r } \right) =  - A\left( \frac{ \del_u\phi }{\del_u r}\right) 
+ \left( \kappa V^\prime(\phi)  - \frac{\del_v\phi}{r} \right). \label{dvfrac}
\end{equation}
Using (\ref{Vbound}), (\ref{2mr_Rbound}) and the fact that $c_0 < \frac{1}{4r_+^2}$, we estimate
\[ A \geq \displaystyle\frac{1- 2 r_+^2 (2c_0 - \epsilon) - 4c_2 b_0 }{4c_2r_1} >  \displaystyle\frac{ r_+^2 \epsilon - 2c_2 b_0 }{2c_2r_1} =: a_0. \]
The constant $a_0$ is positive by our choice of $b_0$.
Also, (\ref{b1}) and (\ref{b2}) imply that
\[ \left|\kappa V^\prime(\phi)  - \frac{\del_v\phi}{r} \right| \leq 
\left(\frac{b_2}{4c_3} + b_1 r_1^{-1}\right)v^{-p} =: a_1 v^{-p} \]
Then for $(u,v) \in \overline{\mathcal{V}}$, integrating (\ref{dvfrac}) along the outgoing null ray $\{ u \} \times [v_1, v]$ yields
\begin{eqnarray} \hspace*{.3in} \left( \frac{\del_u \phi }{ \del_u r } \right)(u,v) & = & e^{-\int_{v_1}^v  A(u,\bar{v}) \,d\bar{v} } \left( \frac{\del_u \phi }{ \del_u r } \right)(u,v_1) \label{dvfracint}\\
&& \hspace*{.2in} + \int_{v_1}^v e^{-\int_{\tilde{v}}^v A(u,\bar{v}) \, d\bar{v}}\left(\kappa V^\prime(\phi)  - \frac{\del_v\phi}{r} \right)(u,\tilde{v}) \,d\tilde{v},   \nonumber
\end{eqnarray}
so
\begin{eqnarray*} \left| \frac{\del_u \phi }{ \del_u r } \right|(u,v) & \leq & e^{-\int_{v_1}^v  A(u,\bar{v}) \,d\bar{v} } \left| \frac{\del_u \phi }{ \del_u r } \right|(u,v_1) \\
&& \hspace*{.2in} + \int_{v_1}^v e^{-\int_{\tilde{v}}^v A(u,\bar{v}) \, d\bar{v}}\left|\kappa V^\prime(\phi)  - \frac{\del_v\phi}{r} \right|(u,\tilde{v}) \,d\tilde{v} \\
& \leq & \textstyle \frac 12 b_3 v_1^{-p} e^{-\int_{v_1}^v  a_0 \,d\bar{v} }  + \displaystyle\int_{v_1}^v a_1 \tilde{v}^{-p} e^{-\int_{\tilde{v}}^v a_0 \, d\bar{v}} \,d\tilde{v} \\
& \leq & \textstyle \frac 12 b_3 v_1^{-p} e^{a_0(v_1 - v)}  + a_1 e^{-a_0 v} \displaystyle\int_{v_1}^v  \tilde{v}^{-p} e^{a_0 \tilde{v}} \,d\tilde{v}.
\end{eqnarray*} 
Integrating the second term by parts, we have:
\begin{eqnarray*}  a_1 e^{-a_0 v} \int_{v_1}^v  \tilde{v}^{-p} e^{a_0 \tilde{v}} \,d\tilde{v} & = &    
a_1 e^{-a_0 v}  \bigg( a_0^{-1} v^{-p} e^{a_0 v} - a_0^{-1} v_1^{-p} e^{a_0 v_1}   \\
&&  \hspace*{.15in} +  \int_{v_1}^v  a_0^{-1} p \tilde{v}^{-p-1} e^{a_0 \tilde v} \,d\tilde{v} \bigg) \\
& \leq &  a_1  a_0^{-1} v^{-p}   + p a_1 a_0^{-1} e^{-a_0 v} \int_{v_1}^v   \tilde{v}^{-p-1} e^{a_0 \tilde v} \,d\tilde{v}. 
\end{eqnarray*}
Furthermore, we estimate that
\begin{eqnarray*}\int_{v_1}^v   \tilde{v}^{-p-1} e^{a_0 \tilde v} \,d\tilde{v} & = &
\int_{v-\frac{p}{a_0}\log v}^v   \tilde{v}^{-p-1} e^{a_0 \tilde v} \,d\tilde{v} + \int_{v_1}^{v-\frac{p}{a_0}\log v}   \tilde{v}^{-p-1} e^{a_0 \tilde v} \,d\tilde{v} \\
& \leq & e^{a_0 v}\int_{v-\frac{p}{a_0}\log v}^v   \tilde{v}^{-p-1} \,d\tilde{v} + \\
&&  + \,\, 
e^{a_0v-p\log v} \int_{v_1}^{v-\frac{p}{a_0}\log v}   \tilde{v}^{-p-1}  \,d\tilde{v} \\
& = & e^{a_0 v}\left( \frac{-v^{-p}}{p} + \frac{(v-\frac{p}{a_0}\log v)^{-p}}{p} \right)   + \\
&& + \,\, e^{a_0 v}v^{-p} \left( \frac{-(v-\frac{p}{a_0}\log v)^{-p}}{p} + \frac{{v_1}^{-p}}{p} \right)   \\
& \leq & p^{-1}e^{a_0v}v^{-p}\left[ \left( -1 + (1-\textstyle\frac{p}{a_0}v^{-1} \log v)^{-p} \right)   + 
 {v_1}^{-p} \right].
\end{eqnarray*}
Putting it all back together yields
\begin{eqnarray*} \hspace*{-2pt} \left| \frac{\del_u \phi }{ \del_u r } \right|(u,v) & \leq & \textstyle \frac 12 b_3 v_1^{-p} e^{a_0(v_1 - v)} +  a_1  a_0^{-1} v^{-p}   + (p a_1 a_0^{-1} e^{-a_0 v}) \cdot\\
&&  \cdot
\left(p^{-1}e^{a_0v}v^{-p}\left[ \left( -1 + (1-\textstyle\frac{p}{a_0}v^{-1} \log v)^{-p} \right)   + 
 {v_1}^{-p} \right]\right) \\
& \leq & \textstyle \frac 12 b_3 v_1^{-p} e^{a_0(v_1 - v)} +  a_1  a_0^{-1} v^{-p} \left( (1-\textstyle\frac{p}{a_0}v^{-1} \log v)^{-p}   +  {v_1}^{-p}   \right)\\
& < & \left( \textstyle \frac 12 b_3 {v^p}{v_1}^{-p} e^{a_0(v_1 - v)} +  \textstyle\frac{1}{2} b_3  \right) v^{-p}\\
& < & b_3 v^{-p},
\end{eqnarray*}
where in the second to last line we have used the monotonicity of $v^{-1} \log v$ together with the definition of $b_3$ and the fact that $r_1 < r_+$, and in the last line we have used the fact that $v^p e^{-a_0 v}$  decreases monotonically for $v > p a_0^{-1}$, a lower bound which is guaranteed by (\ref{v13}) and our choice of $v_1$ (assuming without loss of generality that $v_1 \geq e$).  Thus we have retrieved (\ref{b3}) in $\overline{\mathcal{V}}$.

\medskip

For (\ref{Vbound}), we compute as in (\ref{intV}) that for $(u,v) \in \overline{\mathcal{V}}$,
\[ V(\phi(u,v)) \leq V(\phi(u, v_1)) + \int_{v_1}^v |\del_v(V(\phi))(u,\tilde{v})|\, d\tilde{v} < c_0 - \textstyle\frac 12 \epsilon + \epsilon^\prime < 2c_0 - \epsilon,\]
where the last inequality follows from our choice of $\epsilon^\prime$.  Thus we have retrieved (\ref{Vbound}) in $\overline{\mathcal{V}}$.

As discussed above, we can now conclude that  $\mathcal{C}^\prime_{out} \subset \mathcal{V}$ and hence that $\mathcal{W} \cap \mathcal{V} \neq \emptyset$.
We now turn to improving inequalities (\ref{b0})-(\ref{b2}) and (\ref{c2})  in $\overline{\mathcal{W} \cap \mathcal{V}}$.

\medskip

For (\ref{b0}), note that equation (\ref{ruv}) and inequalities (\ref{Vbound}) and (\ref{b0}) imply that 
\begin{eqnarray*}
\del^2_{uv}r 
& \leq & \textstyle\frac{1}{4} \Omega^2 r^{-1}  \left( 2 r_+^2 (2c_0 - \epsilon)  + 4 c_2 b_0 -1 \right) \\
& \leq &  \textstyle\frac{1}{4} \Omega^2 r^{-1}  \left(  4 c_2 b_0 - 2 r_+^2 \epsilon    \right) \\
& \leq & 0
\end{eqnarray*}
in $\overline{\mathcal{V}}$.  Since $\del_v r(0,v) < \eta(v) \leq \eta(v_1) = b_0$ for all $v\geq v_1$, this yields (\ref{b0}).

\medskip

Next we turn to (\ref{b1}). Rearranging (\ref{phiuv}), we have
\begin{eqnarray*}  \del^2_{uv}\phi 
& = & (\del_u r)\left( \kappa  V^\prime(\phi) - \frac{\del_u \phi}{\del_u r}(\del_v  \log r) - \frac{\del_v\phi}{r} \right), 
\end{eqnarray*}
so for $(u,v) \in \overline{\mathcal{W} \cap \mathcal{V}}$, using inequalities (\ref{b0})-(\ref{b3}) and the fact that $r \geq r_+ - \delta$ implies 
\[
|\del^2_{uv}\phi (u,v) | \leq -(\del_u  r(u,v))\left( \frac{b_2}{4c_3} + \frac{b_0 b_3 + b_1}{r_+-\delta}  \right) v^{-p}. 
\]
Thus
\begin{eqnarray*} |\del_v \phi(u,v) | & \leq & |\del_v \phi(0,v) | + \int_0^u |\del^2_{uv} \phi (\tilde{u},v)|\,d\tilde{u}\\
& \leq & |\del_v \phi(0,v) | - \left( \frac{b_2}{4c_3} + \frac{b_0 b_3 + b_1}{r_+-\delta}  \right) v^{-p} \int_0^u \del_u  r(\tilde{u},v) \,d\tilde{u}\\
& < & \textstyle\frac{1}{2} b_1 v^{-p} + \displaystyle\delta\left( \frac{b_2}{4c_3} + \frac{2b_0 b_3 + 2b_1}{r_+}  \right) v^{-p} \\
& < &  b_1 v^{-p},  
\end{eqnarray*}  
where in the second-to-last and last lines we have used our choice of $\delta$.  Thus we have obtained (\ref{b1}) in $\overline{\mathcal{W} \cap \mathcal{V}}$.

\medskip

Next we retrieve (\ref{b2}).  First observe that for $(u,v) \in \overline{\mathcal{W} \cap \mathcal{V}}$,
\[ \left| \int_0^u (\del_u \phi)(\tilde{u},v) \,d\tilde{u} \right| 
\leq  - b_3 v^{-p} \int_0^u \del_u r(\tilde{u},v) \,d\tilde{u} 
\leq \, b_3 v_1^{-p} \delta \,
< \delta_0,
\]
where we have used (\ref{b3}) and our choice of $\delta$.  
Thus $\phi(u,v) \in (\phi_0 - \delta_0, \phi_1 + \delta_0)$, so in particular,
$|V^{\prime\prime}(\phi(u,v))| \leq B$ for all $(u,v) \in \overline{\mathcal{W} \cap \mathcal{V}}$.
Using (\ref{b3}) once more, we have:
\begin{eqnarray*} |V^\prime(\phi(u,v))| & \leq & |V^\prime(\phi(0,v))| + \int^u_0 |V^{\prime\prime}(\phi)||\del_u \phi| (\tilde{u}, v)\, d\tilde{u} \\
& \leq & \textstyle\frac 12 b_2 v^{-p} - B b_3v^{-p} \displaystyle\int_0^u \del_u r (\tilde{u},v)\, d\tilde{u} \\
& \leq & \textstyle\frac 12  b_2 v^{-p} + B b_3\delta v^{-p} \\
& < & b_2v^{-p},  
\end{eqnarray*}
where in the last line we have again used our choice of $\delta$.

\medskip

It remains only to retrieve (\ref{c2}). Note that from (\ref{eq1}) we have
\[ \del_u(- \Omega^{-2} \del_u r) = r \Omega^{-2} (\del_u\phi)^2, \]
and combining (\ref{c2}) and (\ref{b3}) yields 
\[ \Omega^{-2} (\del_u\phi)^2 \leq b_3^2 v^{-2p}\, \Omega^{-2} (\del_u r )^2 < - b_3^2 c_2 v^{-2p} (\del_u r). \]
Integrating along an ingoing null ray and using (\ref{c2c3choice}), we have
\begin{eqnarray*}  (-\Omega^{-2} \del_u r) (u,v) & \leq & 
(-\Omega^{-2} \del_u r) (0,v) - \int_0^u b_3^2 c_2 v^{-2p} r (\del_u r) \,d\tilde{u} \\
& \leq & \textstyle\frac{1}{2}c_2 + b_3^2 c_2 v^{-2p} r_+ \delta \\
& < & c_2, 
\end{eqnarray*}
 once more using the choice of $\delta$.

\medskip

Thus inequalities (\ref{b0})-(\ref{Vbound}) hold in all of $\overline{\mathcal{W} \cap \mathcal{V}}$, which implies that the boundary of $\mathcal{W} \cap \mathcal{V}$ relative to $\mathcal{W} \cap \mathcal{R}$ is empty.  Thus $\mathcal{W} \cap \mathcal{V} = \mathcal{W} \cap \mathcal{R}$, so in particular, $\mathcal{W} \cap \mathcal{R} \subset \mathcal{V}$ and the theorem follows. 
\end{proof}

\smallskip

%%%%%%%%%%%%%%%%%%%%%%%%%%%%%% PROOF OF HIGGS_M %%%%%%%%%%%%%%%%%%%%%%%%%%%%%%%%%%%%%%%%%%%%%
%		     (the M stands for "Monotonicity")

\begin{proof}[Proof of Theorem \ref{HiggsM}]
Consider the maximal development $\mathcal{G}(u_0, v_0)$ of the given initial data, and define a region $\mathcal{V}_{0}$ as the set of all $(u,v) \in  \mathcal{G}(u_0, v_0)$ \linebreak such that the following ten inequalities hold for all  $(\tilde{u}, \tilde{v}) \in J^- (u,v)$:  
\begin{eqnarray} 
\del_v\phi & < & 0 \label{dvphineg} \\
\del_u\phi & < & 0 \label{duphineg} \\
V^\prime(\phi) & < & c_4 |\del_v \phi | \label{Vprime}\\
V^\prime (\phi) - 4\sqrt{c_1}c_2 (\del_v \log r) - 4c_3r_+^{-1}(\del_v \phi) & > & 0, \label{bigineq} \\
\del_v r & < & \epsilon \label{ep}\\
V(\phi) & < & \epsilon^\prime \label{ep2} \\
|\del_v\phi| & < & \epsilon^{\prime\prime} \label{ep3} \\
| \del_u\phi | & < & \sqrt{ c_1} |\del_u r| \label{c1} \\
-(\del_u r) \Omega^{-2} & < &  c_2 \label{c2M} \\
\del_v r & > & 0.  \label{regM} 
\end{eqnarray}
Clearly $\mathcal{V}_0$ is open in $\mathcal{G}(u_0, v_0)$.  Consequently, our assumptions on the initial data imply that $\mathcal{V}_0$ must contain some neighborhood of $(0, v_0)$ in $\mathcal{G}(u_0, v_0)$, so by shrinking $u_0$ as necessary, we may in fact assume that they all hold along $\mathcal{C}_{in}$.  Since all of the inequalities except (\ref{c1}) are known to hold on $\mathcal{C}_{out}$, our first step will be to retrieve (\ref{c1}) in $\overline{\mathcal{V}_0}$, where the set closure is taken relative to $\mathcal{G}(u_0, v_0)$, i.e. $\overline{\mathcal{V}_0} = \overline{\mathcal{V}_0} \cap \mathcal{G}(u_0, v_0)$.  Then by a continuity argument, we can conclude that $\mathcal{C}_{in} \cup \mathcal{C}_{out} \subset \mathcal{V}_0$.

As in the proof of Theorem \ref{HiggsR}, recall that quantity $\kappa$ is given by
\[ \kappa = -\textstyle\frac 14 \Omega^2 (\del_u r)^{-1} = \displaystyle\frac{(\del_v r)}{1-\frac{2m}{r}}.\]
From (\ref{c2M}), we have $\kappa \geq \frac{1}{4c_2}$ in $\overline{\mathcal{V}_0}$, and since $\del_u \kappa \leq 0$ by equation (\ref{eq1}), (\ref{c2c3choiceM})
implies
$ \kappa \leq \frac{1}{4c_3} $
in all of $\mathcal{G}(u_0, v_0)$.
Also, from (\ref{ep}) we have that
\begin{equation} \textstyle(1-\frac{2m}{r}) = (\del_v r) \kappa^{-1} \leq 4 c_2 \epsilon \label{2mrep}\end{equation}
in $\overline{\mathcal{V}_0}$.
Let $r_0 = r(0, v_0)$ and observe that $\del_v r > 0$ implies that $r \geq r_0$ on all of $\mathcal{C}_{out}$.  

Now, equation (\ref{dvfracint}) may be derived exactly as in the proof of Theorem \ref{HiggsR}, namely
\begin{eqnarray*} \left( \frac{\del_u \phi }{ \del_u r } \right)(u,v) & = & e^{-\int_{v_0}^v  A(u,\tilde{v}) \,d\tilde{v} } \left( \frac{\del_u \phi }{ \del_u r } \right)(u,v_0) \\
&& \hspace*{.2in} + \int_{v_0}^v e^{-\int_{v^\prime}^v A(u,\tilde{v}) \, d\tilde{v}}\left(\kappa V^\prime(\phi)  - \frac{\del_v\phi}{r} \right)(u,v^\prime) \,dv^\prime,
\end{eqnarray*}
where
\[ 
A := \kappa r^{-1} \left[ (1- 2 r^2 V(\phi)) - (1-\textstyle\frac{2m}{r}) \right] + (\del_v \log r).
\]
Using the above bounds, (\ref{ep2}), and (\ref{regM}), in $\overline{\mathcal{V}_0}$ we estimate
\[ A \geq \displaystyle\frac{1- 2 r_+^2 \epsilon^\prime - 4c_2 \epsilon }{4c_2r_0} =: a_0. \]
The constant $a_0$ is positive by our choices of $\epsilon$ and $\epsilon^\prime$.
Also, (\ref{ep3}) and (\ref{Vprime}) imply that
\[  \kappa V^\prime(\phi)  - \frac{\del_v\phi}{r}  \leq \left(\frac{c_4}{4c_3}+ \frac{1}{r_0}\right)\epsilon^{\prime\prime} =: a_1. \]
Then applying these bounds and using (\ref{c1choice}), for $(u,v) \in \overline{\mathcal{V}_0}$ we have 
\begin{eqnarray*}  \left(\frac{\del_u \phi }{ \del_u r }\right)(u,v)  
& < & \sqrt{c_1} \,e^{-a_0(v-v_0)}  + a_1 \int_{v_0}^v e^{-a_0(v-v^\prime)} \,dv^\prime  \\
& = & \sqrt{c_1} \, e^{-a_0(v-v_0)}  + 
a_1 a_0^{-1}\left(1 - e^{-a_0(v-v_0)} \right) \\
& = & e^{-a_0(v-v_0)} \left( \sqrt{c_1} - a_1 a_0^{-1} \right) + a_1 a_0^{-1}.
\end{eqnarray*}
Our choices of $\epsilon$, $\epsilon^\prime$ and $\epsilon^{\prime\prime}$ imply that $\sqrt{c_1} - a_1 a_0^{-1} > 0$, so
for $v \geq v_0$, we have 
\[ \left(\frac{\del_u \phi }{ \del_u r }\right)(u,v) <   \left( \sqrt{c_1} - a_1 a_0^{-1} \right) +  a_1 a_0^{-1} = \sqrt{c_1}. \]
Thus (\ref{c1}) holds in all of $\overline{\mathcal{V}_0}$, so in particular, $\mathcal{C}_{out} \subset \mathcal{V}_0$.

\medskip

Our next step is to choose a suitably small $\delta > 0$ to use in defining $\mathcal{W}$:  we let 
\[ \delta < \min\left\{ \frac{\delta_0}{\sqrt{c_1}}, \,\frac{\epsilon^\prime r_+}{8\sqrt{c_1}c_3\epsilon^{\prime\prime}}, 
\,\, r_+\!\left(1-2^{-\left(\frac{c_4 r_+}{4c_3} + 1\right)^{-1}} \right),  
\,\frac{1}{2c_1 r_+}
\right\}\]
and set $\mathcal{W} = \mathcal{W}(\delta) = \{ (u,v) \in \mathcal{G}(u_0, v_0) \,|\, r(u,v) \geq r_+ - \delta \}$.
Now, $r \nearrow r_+$ along $\mathcal{C}_{out}$, so there must exist some $v_1 \geq v_0$ such that $\mathcal{W}$ contains an open neighborhood of the ray $\{ 0 \} \times [v_1, \infty)$.  Also, since $\mathcal{V}_0$ and $\mathcal{W}$ each contain some neighborhood of the point $(0, v_1)$, we can find $0 < u_1 \leq u_0$ such that $[0, u_1] \times \{ v_1 \} \subset \mathcal{V}_0\cap \mathcal{W}$.  Set $\mathcal{C}_{in}^\prime = [0, u_1] \times \{ v_1 \}$ and $\mathcal{C}_{out}^\prime = \{0 \} \times [v_1, \infty)$.
Henceforth we restrict our attention to the subset $\mathcal{G}(u_1, v_1) = K(u_1, v_1) \cap \mathcal{G}(u_0, v_0)$, that is, the maximal development of the induced data on 
$\mathcal{C}_{in}^\prime \cup \mathcal{C}_{out}^\prime$. 

The proof now proceeds by a bootstrap argument.  
Let $\mathcal{V}$ be the set of all points   $(u,v) \in \mathcal{G}(u_1, v_1)$ such that
$(\tilde{u}, \tilde{v}) \in \mathcal{V}_0 \cap \mathcal{W}$ for all $(\tilde{u}, \tilde{v}) \in J^-(u,v)  \cap  \mathcal{G}(u_1, v_1)$.
Clearly $\mathcal{V} \subset \mathcal{W} \cap \mathcal{R}$.
By construction, we have $\mathcal{C}_{in}^\prime \cup \mathcal{C}_{out}^\prime \subset \mathcal{V}$.  We will retrieve inequalities (\ref{dvphineg})-(\ref{c2M}) in $\overline{\mathcal{V}}$ and consequently conclude that $\mathcal{V} = \mathcal{W} \cap \mathcal{R} \cap \mathcal{G}(u_1, v_1)$.  At that point we can easily extract the hypotheses of Theorem \ref{mainthm}.

\medskip

We proceed through the ten inequalities in order,  beginning with (\ref{dvphineg}) and (\ref{duphineg}).  Rearranging  equation (\ref{sfequv}) and applying (\ref{c1}) and our bounds for $\kappa$ yields
\begin{eqnarray*}  \del^2_{uv}\phi 
& = & -\textstyle\frac{1}{4}\Omega^2 \displaystyle \left( V^\prime(\phi) - \frac{\del_u \phi}{\del_u r}(\del_v  \log r)\kappa^{-1} - \frac{\del_v\phi}{r\kappa} \right) \\
& \leq & -\textstyle\frac{1}{4}\Omega^2 \displaystyle \left( V^\prime(\phi) - 4\sqrt{c_1}c_2(\del_v  \log r) - 4c_3r_+^{-1}(\del_v\phi) \right),
\end{eqnarray*}
so (\ref{bigineq}) now implies that
\begin{equation}\label{phiuvineq} \del^2_{uv}\phi \leq 0
\end{equation}
in $\overline{\mathcal{V}}$.
Thus since (\ref{dvphineg}) and (\ref{duphineg}) hold along $\mathcal{C}^\prime_{out}$ and $\mathcal{C}_{in}^\prime$, respectively, they must hold in $\overline{\mathcal{V}}$ as well.  

\medskip

Now we turn to (\ref{Vprime}) and we compute that
\[ \del_u \left( \frac{V^\prime(\phi)}{\del_v \phi} \right) = \frac{V^{\prime\prime}(\phi)(\del_u \phi)(\del_v \phi) - V^\prime(\phi)(\del^2_{uv} \phi)}{(\del_v \phi)^2} \geq - \frac{V^\prime(\phi) (\del^2_{uv} \phi)}{(\del_v \phi)^2}.
\]
Suppose $(u,v) \in \overline{\mathcal{V}}$.  
If $V^\prime(\phi(u,v)) \leq 0$, then clearly (\ref{Vprime}) holds at $(u,v)$, since $|\del_v \phi| > 0$.  
If $V^\prime(\phi(u,v)) > 0$, let $u_*$ be the smallest value such that $V^\prime(\phi) \geq 0$ along $[u_*, u) \times \{ v \}$ and integrate the above inequality, noting that the righthand side is positive along this ray by (\ref{phiuvineq}). 
If $u_* = 0$, then by our hypotheses on $\mathcal{C}_{out}$, we have  $V^\prime(\phi(u_*,v))(\del_v \phi(u_*,v))^{-1}  > -c_4$. 
On the other hand, if $u_* > 0$, then by choice of $u_*$,  $V^\prime(\phi(u_*,v))(\del_v \phi(u_*,v))^{-1}  = 0$.  
Thus in either case we have
\[   \left( \frac{V^\prime(\phi)}{\del_v \phi} \right)(u,v) \geq  \left( \frac{V^\prime(\phi)}{\del_v \phi} \right)(u_*, v) > -c_4, \]
thereby obtaining (\ref{Vprime}) in $\overline{\mathcal{V}}$.   

\medskip

Next, before proceeding to (\ref{bigineq}), let us first show that our initial bounds for $V^{\prime\prime}(\phi)$ continue to hold in $\overline{\mathcal{V}}$.  
On one hand, integrating (\ref{c1}) yields
\[ \phi(0,v) - \phi(u,v) \leq  \sqrt{c_1}(r(0,v) - r(u,v)) < \sqrt{c_1}\delta < \delta_0. \]
On the other hand, (\ref{duphineg}) implies
\[ \phi(u,v) \leq \phi(0,v). \]
Thus for $(u,v) \in \overline{\mathcal{V}}$ 
we have $\phi(u,v) \in (\phi_0 - \delta_0, \phi_1)$, and hence 
\[ 0 \leq V^{\prime\prime}(\phi(u,v)) \leq B. \] 

\medskip

For (\ref{bigineq}), we first derive an upper estimate for $\del^2_{uv}r$ in $\overline{\mathcal{V}}$.  Recalling equation (\ref{ruv}), we have
\begin{eqnarray*} \del^2_{uv} r 
& = & \textstyle\frac{1}{4} \Omega^2 r^{-1} \left[ 2 r^2 V(\phi)  + (1-\frac{2m}{r}) -1 \right]  \\
& \leq & -\kappa (\del_u r) r^{-1} \left[ 2 r_+^2 \epsilon^\prime  + 4c_2\epsilon -1 \right]  \\
& \leq &  \frac{\del_u r}{4c_2r_+}  \left[ 1- 2 r_+^2 \epsilon^\prime  - 4c_2\epsilon  \right] ,
\end{eqnarray*}
where we have used (\ref{ep2}), (\ref{2mrep}), and $\kappa \geq \frac{1}{4c_2}$. 
Setting
\[ a_2 :=  \frac{1}{4c_2r_+}  \left[ 1- 2 r_+^2 \epsilon^\prime  - 4c_2\epsilon  \right] > 0,\]
we thus have 
\begin{equation}\label{ruvbound} \del^2_{uv} r \leq a_2 (\del_u r). 
\end{equation}
Consequently, differentiating the lefthand side of (\ref{bigineq}) and using inequalities (\ref{c1}), (\ref{duphineg}), (\ref{phiuvineq}), (\ref{ep}), and (\ref{ruvbound})  yields
\begin{eqnarray*}
&& \hspace*{-1.1in}  \del_u \!\left( V^\prime(\phi) - 4\sqrt{c_1}c_2(\del_v \log r) - 4c_3r_+^{-1}(\del_v\phi) \right)  \\ 
& \hspace*{.8in} = & V^{\prime\prime}(\phi)(\del_u \phi) - 4\sqrt{c_1}c_2(\del^2_{uv} \log r) - 4c_3r_+^{-1}(\del^2_{uv}\phi)  \\ 
& \hspace*{.8in} \geq &  \sqrt{c_1} B (\del_u r) - 4\sqrt{c_1}c_2 r^{-1} \del^2_{uv}r  + 4\sqrt{c_1}c_2 r^{-2}(\del_u r)(\del_v r) \\ 
& \hspace*{.8in} \geq &  \sqrt{c_1} B (\del_u r) - 4\sqrt{c_1}c_2  a_2 r_+^{-1} (\del_u r) + 4\sqrt{c_1}c_2 r_+^{-2} \epsilon (\del_u r) \\ 
& \hspace*{.8in} = &  \sqrt{c_1} (\del_u r) \left(B  - 4c_2  a_2 r_+^{-1}  + 4c_2 r_+^{-2} \epsilon\right) \\ 
& \hspace*{.8in} = &  \sqrt{c_1} (\del_u r) \left(B  - r_+^{-2}  + 2\epsilon^\prime + 8c_2 r_+^{-2} \epsilon\right) \\
& \hspace*{.8in} > &  0,
\end{eqnarray*}
where the last line follows from the choices of $\epsilon$ and $\epsilon^\prime$.
Thus we have retrieved (\ref{bigineq}) in $\overline{\mathcal{V}}$.

\medskip

Inequality (\ref{ep}) follows immediately from (\ref{ruvbound}), since the latter implies that $\del^2_{uv} r < 0$.
 
\medskip

For (\ref{ep2}), observe that from (\ref{bigineq}), we have
\[  V^\prime (\phi) \geq  4\sqrt{c_1}c_2 (\del_v \log r) + 4c_3r_+^{-1}(\del_v \phi) \geq 4c_3r_+^{-1}(\del_v \phi).   \]
Multiplying through by $\del_u \phi$ and using (\ref{duphineg}), (\ref{c1}), and (\ref{ep3}) yields
\begin{eqnarray*}  \del_u (V(\phi)) & \leq & 4c_3r_+^{-1} | \del_v \phi || \del_u \phi | \\
& \leq & -4\sqrt{c_1}c_3r_+^{-1} \epsilon^{\prime\prime} ( \del_u r ).
\end{eqnarray*}
Integrating and using the assumption that $V(\phi) < \frac{1}{2} \epsilon^\prime$ on $\mathcal{C}_{out}$ then gives
\begin{eqnarray*} V(\phi)(u,v) 
& \leq & \textstyle\frac{1}{2} \epsilon^{\prime} + 4\sqrt{c_1}c_3r_+^{-1} \epsilon^{\prime\prime} \delta \\
& < & \epsilon^{\prime}
\end{eqnarray*}
by our choice of $\delta$.
 
\medskip

Next we turn to (\ref{ep3}).  Using equation (\ref{sfequv}) and inequalities (\ref{dvphineg})-(\ref{Vprime}), we have
\begin{eqnarray*} \del_u \log |\del_v \phi | & = & (\del^2_{uv} \phi )(\del_v \phi)^{-1} \\
& = & -\textstyle\frac{1}{4}\Omega^{2} V^\prime(\phi)(\del_v \phi)^{-1} - (\del_u\phi)(\del_v \phi)^{-1}(\del_v \log r) - (\del_u \log r) \\
& \leq & -c_4\kappa (\del_u r) - \del_u \log r \\
& \leq & -\del_u \log r \left(\frac{c_4 r_+}{4c_3}  + 1 \right),
\end{eqnarray*}
so integrating yields
\begin{eqnarray*} |\del_v \phi (u,v) | 
& \leq & |\del_v \phi (0,v) |\left( \frac{r_+}{r_+ - \delta} \right)^{\frac{c_4r_+}{4c_3} + 1} \\
& < & \textstyle\frac{1}{2} \epsilon^{\prime\prime} \displaystyle\left( \frac{r_+}{r_+ - \delta} \right)^{\frac{c_4r_+}{4c_3} + 1} \\
& < & \epsilon^{\prime\prime},
\end{eqnarray*}
where in the last line we have again used our choice of $\delta$. Thus (\ref{ep3}) holds in $\overline{\mathcal{V}}$.

\medskip

We have already shown that (\ref{c1}) holds in $\overline{\mathcal{V}_0}$, so naturally it holds in $\overline{\mathcal{V}}$ as well.

\medskip

Lastly we retrieve (\ref{c2M}).   Note that from (\ref{eq1}) we have
\[ \del_u(- \Omega^{-2} \del_u r) = r \Omega^{-2} (\del_u\phi)^2, \]
and combining (\ref{c2M}) and (\ref{c1}) yields 
\[ \Omega^{-2} (\del_u\phi)^2 \leq c_1 \Omega^{-2} (\del_u r )^2 < - c_1 c_2 (\del_u r). \]
Integrating along an ingoing null ray and using (\ref{c2M}) again, we have
\begin{eqnarray*}  (-\Omega^{-2} \del_u r) (u,v) & \leq & 
(-\Omega^{-2} \del_u r) (0,v) - \int_0^u c_1 c_2 r (\del_u r) \,d\tilde{u} \\
& \leq & \textstyle\frac{1}{2}c_2 + c_1 c_2 r_+ \delta \\
& < & c_2,
\end{eqnarray*}
once more using our choice of $\delta$ in the last line.

\medskip

The bootstrap is now completed; we have shown that inequalities (\ref{dvphineg})-(\ref{c2M}) hold in all of $\mathcal{W} \cap \mathcal{R}$, and hence that
$\mathcal{W} \cap \mathcal{R} \subset \mathcal{V}$.  It remains to show that the hypotheses of Theorem \ref{mainthm} hold in this region.

For ${\rm{\Aprimecond}}^\prime$, we note that by (\ref{ep2}), $\Omega^{-2}T_{uv} = \frac 12 V(\phi) < \frac 12 \epsilon^\prime < c_0 < \frac 14 (r_+)^{-2}$.  Conditions \Bcondone\, and \Ccond\, are immediate by (\ref{c1}) and (\ref{c2M}), respectively.  
For the second part of condition \Bcondtwo, we use (\ref{ep2}) and the nonnegativity of $V(\phi)$ to estimate that 
\[ \int_{v_0}^v \del_v (\Omega^{-2}T_{uv}) \, d\tilde{v} = \int_{v_0}^v \del_v (\textstyle\frac 12 V(\phi)) \, d\tilde{v} = \frac 12 \left[ V(\phi(u,v)) - V(\phi(u, v_0))\right] < \frac 12 \epsilon^\prime \]
and then observe that $\frac 12 \epsilon^\prime < \frac{1}{4r_+^2} - c_0$.
For the first part of \Bcondtwo,
recall that one of our hypotheses was that either $V^\prime(\phi) \leq 0$ along $\mathcal{C}_{out}$ or $|\phi_0| < \infty$.  
In the former case,  we may differentiate $V^\prime(\phi)$ and use (\ref{duphineg}) and the nonnegativity of $V^{\prime\prime}$ to obtain
\[ \del_u (V^\prime(\phi)) = V^{\prime\prime}(\phi) (\del_u\phi) \leq 0.\]
Thus $V^\prime(\phi) \leq 0$ in all of $\overline{\mathcal{V}}$, so combining this with inequality (\ref{dvphineg}),  we have $| \del_v (\frac 12 V(\phi) | = \del_v (\frac 12 V(\phi))$, and the first part of \Bcondtwo\, now follows from the second.  
In the latter case, we have $\phi_0 - \delta_0 < \phi(u,v) < \phi_1$  for all $(u,v) \in \mathcal{V}$, where we are now assuming that $|\phi_0| < \infty$.  Since $\phi$ decreases along $\mathcal{C}_{out}$, $|\phi_1| < \infty$ as well. 
Thus, using (\ref{Vprime}), (\ref{dvphineg}), and (\ref{ep3}), we have
\begin{eqnarray*} \int_{v_0}^v | \del_v (\textstyle\frac 12 V(\phi))| \, d\tilde{v} & = & \displaystyle\int_{v_0}^v \textstyle\frac 12 |V^\prime(\phi)||\del_v \phi| \, d\tilde{v} \\
& < & - \displaystyle\int_{v_0}^v \textstyle\frac 12 c_4 \epsilon^{\prime\prime} (\del_v \phi) \, d\tilde{v} \\
& < & \textstyle\frac 12 c_4 \epsilon^{\prime\prime} (\phi(u, v_0) - \phi(u, v)) \\
& < & \textstyle\frac 12 c_4 \epsilon^{\prime\prime} (\phi_1 - \phi_0 + \delta_0) \\
& < & \infty.
\end{eqnarray*}
Thus condition \Bcondtwo\, holds in either case.  
The verification of assumptions \DEC-\extprinc\, is identical to that in the proof of Theorem \ref{HiggsR}.
\end{proof}

\bibliography{mybib}{}

\providecommand{\bysame}{\leavevmode\hbox to3em{\hrulefill}\thinspace}
\providecommand{\MR}{\relax\ifhmode\unskip\space\fi MR }
\providecommand{\MRhref}[2]{%
  \href{http://www.ams.org/mathscinet-getitem?mr=#1}{#2}
}
\providecommand{\href}[2]{#2}
\begin{thebibliography}{10}

\bibitem{AMS}
Lars Andersson, Marc Mars, and Walter Simon, \emph{Local existence of dynamical
  and trapping horizons}, Physical Review Letters \textbf{95} (2005), no.~11,
  111102.

\bibitem{AM}
Lars Andersson and Jan Metzger, \emph{Curvature estimates for stable marginally
  trapped surfaces},  (2005), gr-qc/0512106.

\bibitem{AG}
Abhay Ashtekar and Gregory~J. Galloway, \emph{Some uniqueness results for
  dynamical horizons}, Adv. Theor. Math. Phys. \textbf{9} (2005), no.~1, 1--30.

\bibitem{A}
Abhay Ashtekar and Badri Krishnan, \emph{Isolated and dynamical horizons and
  their applications}, Living Reviews in Relativity \textbf{7} (2004), no.~10.

\bibitem{B2}
Luca Baiotti, Ian Hawke, Pedro~J. Montero, Frank Loffler, Luciano Rezzolla,
  Nikolaos Stergioulas, Jose~A. Font, and Ed~Seidel, \emph{Three-dimensional
  relativistic simulations of rotating neutron-star collapse to a {K}err black
  hole}, Physical Review D (Particles, Fields, Gravitation, and Cosmology)
  \textbf{71} (2005), no.~2, 024035.

\bibitem{B}
Ivan Booth, Lionel Brits, Jose~A. Gonzalez, and Chris Van Den~Broeck,
  \emph{Marginally trapped tubes and dynamical horizons}, Classical Quantum
  Gravity \textbf{23} (2006), no.~2, 413--439.

\bibitem{C2}
Demetrios Christodoulou, \emph{Bounded variation solutions of the spherically
  symmetric {E}instein-scalar field equations}, Comm. Pure Appl. Math.
  \textbf{46} (1993), no.~8, 1131--1220.

\bibitem{C}
\bysame, \emph{On the global initial value problem and the issue of
  singularities}, Classical Quantum Gravity \textbf{16} (1999), no.~12A,
  A23--A35.

\bibitem{D1}
Mihalis Dafermos, \emph{The interior of charged black holes and the problem of
  uniqueness in general relativity}, Comm. Pure Appl. Math. \textbf{58} (2005),
  no.~4, 445--504.

\bibitem{D3}
\bysame, \emph{On naked singularities and the collapse of self-gravitating
  {H}iggs fields}, Adv. Theor. Math. Phys. \textbf{9} (2005), no.~4, 575--591.

\bibitem{D2}
\bysame, \emph{Spherically symmetric spacetimes with a trapped surface},
  Classical Quantum Gravity \textbf{22} (2005), no.~11, 2221--2232.

\bibitem{DR2}
Mihalis Dafermos and Alan~D. Rendall, \emph{An extension principle for the
  {E}instein-{V}lasov system in spherical symmetry}, Ann. Henri Poincar\'e
  \textbf{6} (2005), no.~6, 1137--1155.

\bibitem{DR}
\bysame, \emph{Strong cosmic censorship for surface-symmetric cosmological
  spacetimes with collisionless matter},  (2007), gr-qc/0701034.

\bibitem{H}
Sean~A. Hayward, \emph{General laws of black-hole dynamics}, Phys. Rev. D (3)
  \textbf{49} (1994), no.~12, 6467--6474.

\bibitem{P}
Richard~H. Price, \emph{Nonspherical perturbations of relativistic
  gravitational collapse. {I}. {S}calar and gravitational perturbations}, Phys.
  Rev. D (3) \textbf{5} (1972), 2419--2438.

\bibitem{SKB}
Erik Schnetter, Badri Krishnan, and Florian Beyer, \emph{Introduction to
  dynamical horizons in numerical relativity}, Physical Review D (Particles,
  Fields, Gravitation, and Cosmology) \textbf{74} (2006), no.~2, 024028.

\bibitem{WL}
B.~Waugh and Kayll Lake, \emph{Double-null coordinates for the {V}aidya
  metric}, Phys. Rev. D \textbf{34} (1986), no.~10, 2978--2984.

\end{thebibliography}
\bibliographystyle{amsplain}

\end{document}